\documentclass[onecolumn,aps,prd,preprintnumbers,showpacs,superscriptaddress,nofootinbib,amsmath,amssymb,floats,floatfix,showkeys,notitlepage,longbibliography]{revtex4-1}

\usepackage{orcidlink}
\usepackage{lipsum}
\usepackage{graphicx}
\usepackage{subfigure}
\usepackage{braket}
\usepackage[utf8]{inputenc}
\usepackage{sans}
\usepackage{changes}
\usepackage{hyperref}
\hypersetup{colorlinks=true,linkcolor=blue,urlcolor=blue,citecolor=blue}
\usepackage[toc,page]{appendix}
\usepackage[normalem]{ulem}
\usepackage{adjustbox}
\usepackage{latexsym}
\usepackage{amsmath}
\usepackage{amssymb}
\usepackage{amsfonts}
\usepackage{dcolumn}
\usepackage{bm}
\usepackage{tikz}
\usepackage{bigints}
\usepackage{comment}
\usepackage{array,tabularx,multirow,booktabs}
\usepackage[tracking=true]{microtype}
\SetTracking{}{500}
\SetTracking{encoding={*}, shape=sc}{40}
\UseRawInputEncoding 
\allowdisplaybreaks

\begin{document} \sloppy

\title{
 Black holes immersed in  polytropic scalar field gas}

\author{Y. Sekhmani\orcidlink{0000-0001-7448-4579}}
\email{sekhmaniyassine@gmail.com}
\affiliation{Center for Theoretical Physics, Khazar University, 41 Mehseti Street, Baku, AZ1096, Azerbaijan.}
\affiliation{Centre for Research Impact \& Outcome, Chitkara University Institute of Engineering and Technology, Chitkara University, Rajpura, 140401, Punjab, India.}

\author{S. Zare}
\email{szare@uva.es}
\affiliation{Departamento de F\'{i}sica Te\'{o}rica, At\'{o}ómica y Optica and Laboratory for Disruptive Interdisciplinary Science (LaDIS),
Universidad de Valladolid, 47011 Valladolid, Spain}

\author{L.M. Nieto\orcidlink{0000-0002-2849-2647}}
\email{luismiguel.nieto.calzada@uva.es}
\affiliation{Departamento de F\'{i}sica Te\'{o}rica, At\'{o}ómica y Optica and Laboratory for Disruptive Interdisciplinary Science (LaDIS),
Universidad de Valladolid, 47011 Valladolid, Spain}

\author{H. Hassanabadi\orcidlink{0000-0001-7487-6898}}
\email{hha1349@gmail.com}
\affiliation{Department   of   Physics,   University   of   Hradec   Kr\'{a}lov\'{e}, Rokitansk\'{e}ho   62,   500   03   Hradec   Kr\'{a}lov\'{e},   Czechia}

\author{K. Boshkayev\orcidlink{0000-0002-1385-270X}}
\email{kuantay@mail.ru}
\affiliation{Al-Farabi Kazakh National University, Al-Farabi av. 71, 050040 Almaty, Kazakhstan
}
\affiliation{Institute of Nuclear Physics, Ibragimova, 1, 050032 Almaty, Kazakhstan}
\affiliation{Kazakh-British Technical University, Tole bi str., 59, Almaty, 050000, Kazakhstan}

\date{\today}

\begin{abstract}
By implementing the concept of polytropic structures as a scalar field gas with a dark energy-like behavior, we obtain a static spherically symmetric black hole solution in the framework of general relativity. In this paper, we study the quasinormal modes, the greybody bound process, the shadow behaviors, and the sparsity of black holes with a surrounding polytropic scalar field gas. Using the Wentzel-Kramers-Brillouin approach, we evaluate the impact of a particular set of polytropic parameters $(\xi, A)$ with a fixed setting of the polytropic index $n$ on the oscillation frequency and damping rate of gravitational waves. The results show that the effect of the parameter $\xi$ is much less significant than that of the parameter $A$ on the gravitational waves oscillation frequency and damping rate. Furthermore, the analysis of the greybody factor bounds reveals special insight into the effect of certain parameters where the multipole moments $l$ and the polytropic index $n$ have similar effects, in contrast to the pair of polytropic parameters ($\xi,A$). On the other hand, exploring the sparsity of Hawking radiation is another task that provides a better understanding of the behaviour of the black hole solution. In this respect, the results show that the black hole behaves like blackbody radiation for a sufficiently large entropy. And for $\xi=A=0$, the relevant sparsity acts exactly like the Schwarzschild sparsity. These results provide an insight into the dynamics of black holes with a surrounding polytropic scalar field gas from the analysis of their quasinormal modes, greybody factors, shadow behaviors, energy emission rate and sparsity process. 
Constraints on the associated BH parameters, derived from the Event Horizon Telescope observations of M87* and Sgr A*, indicate that this black hole model stands as a compelling candidate for representing astrophysical black holes.
\end{abstract}

\maketitle

\section{Introduction}
black holes (BHs) are thought of as enormously puzzling compact objects in the universe. The presence of these compact objects offers an effective route for discovering gravitational effects within extremely intense gravitational fields, e.g., the disruption of neighbouring stars and the formation of gigantic jets. Although BHs have posed a major challenge to the observing commons, the theoretical perspective on BHs provides an interesting insight into probing theories of gravity beyond general relativity (GR) \cite{Berti:2015itd,Barack:2018yly,LIGOScientific:2021sio,Zhao:2019kif,Zhang:2019iim,Zhang:2020too,Zhang:2023kzs,Carson:2020ter}. Of course, BHs were originated as mathematical solutions to GR \cite{Schwarzschild:1916uq}. Even so, observations have already been made to prove their existence, notably of gravitational waves (GWs), ushering in a promising prospect for GW astronomy \cite{LIGOScientific:2016aoc}. In a bid to model the nature of GW, several GW signals have been predicted as coalescences of compact bodies, BHs, and/or neutron stars as part of the LIGO/Virgo/KAGRA (LVK) scientific collaboration \cite{LIGOScientific:2018mvr,LIGOScientific:2019lzm,KAGRA:2021vkt}. On the other hand, the insistence on modelling the nature of GW will eventually result in the construction of the most sophisticated ground- and space-based GW detectors \cite{Moore:2014lga,Gong:2021gvw}, such as the Cosmic Explorer, Einstein Telescope, Lisa Telescope, Tian Chen, Taiji, and Desigo. GW, at this stage, will be detected over a broader frequency spectrum and at long distances. This has generated a great deal of enthusiasm for exploring the quasinormal modes (QNMs) of BHs \cite{Berti:2009kk,Berti:2018vdi} and the processes of inspiration with extreme mass ratios \cite{Zhang:2022rfr,Tu:2023xab}. 

QNMs are the characteristic oscillations that arise when BH or neutron stars are disturbed by external effects \cite{KokkotasLRR,NollertCQG1999}. These perturbations set off GWs, which act as cosmic messengers, encoding crucial details about the characteristics of these astrophysical objects \cite{KrivanPRD1997,Rezzolla,KonoplyaPRD2005}. The study of QNMs offers valuable insights into the stability of these compact objects \cite{JaramilloPRX2021}.  Unlike normal modes, which oscillate indefinitely and stay stable, QNMs are distinct in that they have complex frequencies, with the real part indicating the oscillation frequency and the imaginary part reflecting the damping rate \cite{KokkotasLRR,NollertJMP1999,BertiPRD2003}. If the damping rate of a frequency is negative, the mode is stable and will gradually fade away over time. On the other hand, a positive damping rate suggests that the system could behave unstably, potentially leading to extreme phenomena such as gravitational collapse \cite{KonoplyaPRD2006,KonoplyaPRD2011}.

In the light of recent observations by EHT, in particular the brand-new event horizon scale images of M87$^\star$ and Sgr A$^\star$ BHs \cite{EHTApJL2019L1,EHTApJL2019L2,EHTApJL2019L3,EHTApJL2019L4,EHTApJL2019L5,EHTApJL2019L6,EHTApJL2022L12,EHTApJL2022L17,DoSci2019,GRAVITY2022,GRAVITY2020,KocherlakotaPRD2021,VagnozziCQG2023} have aroused particular interest in the community of experimental and theoretical physics. Indeed, these recent event horizon scale images offer a concrete way for testing in a theoretical frame the GR theory and such-inspired models of gravity, such as modified theories of gravity (MOG). For that reason, a wide number of investigations have been conducted to perform a comparative study between the observed image and the theoretical image, providing a comparison in view of the size and shape of the observed image of M87$^\star$ or Sgr A$^\star$. On the other hand, the exploration of the optical appearance of BHs has led to connecting this aspect from within the thermodynamic behaviours by linking the event horizon radius with the photon sphere radius \cite{8,9}. In addition, an attempt to discover a further meaning of the photon sphere relevant to the BH shadow has also been carried out from the QNMs process and provided fruitful insight \cite{10,11}. Practically speaking, from a topological point of view, the orbiting photon spheres involve two topologically kind surfaces, either like a $2D$ torus encompassing an infinite photon sphere for the case of static BHs or like an infinite photon sphere embedding into a two-dimensional deformed surface, which is the scenario for the rotating BHs family endowed with an ergosphere region \cite{12,13,14,15,16,Zahid:2023csk,Khan:2023qsl,Rayimbaev:2022hca,Belhaj:2022kek,Belhaj:2022gcj,Belhaj:2023nhq,Gogoi:2023ntt,Sekhmani:2024fjn,Sekhmani:2024dhc,Al-Badawi:2024jnt}. In the recent past, several studies have been carried out on the subject of shadow processing in GR with a specific set of physical parameters or in different contexts as an extension of GR. In this regard, within the context of Einstein-Maxwell dilaton-axion gravity (EMDA), the associated shadow behaviour is performed in view of the action of suitable parameter spaces such that the dilaton and the spin \cite{18}. Similarly, within the context of $f(T)$ teleparallel gravity, the corresponding BH shadow and chaos bound violation mechanisms have been explored by setting the appropriate parameter space \cite{199}. 

In this study, the polytropic structure is considered as the fundamental block for the matter-source energy-impulsion in the Einstein equations. For that reason, it might be useful to describe polytropic structure as an anisotropic formalism background.  Broadly speaking, a quite distinct classification exists, known as polytropic cosmological gases, which fulfill the anthropic principle. Indeed, polytropic structures are typically characterised by the non-linear equation of state (EoS) such that $p( \rho) =-\xi \rho^{1+\frac{1}{n}}$ whith $n$ is the polytropic index ($1<n< \infty$) \cite{Karami:2009zz,Karami:2012by,Karami:2010px,Karami:2010rb,Banerjee:2024qzl,Aboueisha:2023lad,Cardenas:2024ygf,Jia:2024hpw}. In particular, the corresponding EoS establishes in its form a negative pressure exhibiting similar behaviour to dark energy fluids, where for the set $\frac{1}{3}<\xi<1$ and $n\rightarrow\infty$, the polytropic structure is referred to as the quintessence field solution, i.e., a dark energy candidate. The cosmological constant $\Lambda_0$, on the other hand, is another candidate for dark energy, which results from the condition where $\gamma=1$ and $n\rightarrow\infty$. Nevertheless, the situation of the quintessence according to P. Steinhardt's works \cite{Zlatev:1998tr,Steinhardt:1999nw,Caldwell:1999ew} is represented to be the second candidate of dark energy, which is treated as a scalar field predicting and explaining the accelerating of the universe. 

Approaching the polytropic structure in the essence of gravitational compact objects, such as BHs, and from the point of view of cosmological investigations upon dark energy models constituted a concrete and verifiable framework \cite{Karami:2009zz,Karami:2012by,Karami:2010px,Karami:2010rb,Banerjee:2024qzl,Aboueisha:2023lad,Cardenas:2024ygf,Jia:2024hpw}. In particular, from that point of view, the polytropic structure with the non-linear EoS can be regarded as an alternative way in such cosmological models leading to the polytropic cosmological $\xi$-models. Among several investigations, carrying out an examination in terms of the dark energy with a cold matter scenario by considering the cosmological aspect of the polytropic structure systems \cite{Cardenas:2024ygf}, which is properly leading to intriguing results and novel interpretations. Indeed, this investigation is realised by taking into account one of the important kinds of polytropic structure, nothing more than Chaplygin gas. In concrete terms, the Chaplygin gas here is described with EoS like $p(\rho)=-\xi \rho^{-1}$ with $n=-\frac{1}{2}$ referred to the polytropic index. On the other hand, and similarly to the Chaplygin gas, the Quartessence is another dark fluid configuration unifying in a single component dark energy and dark matter \cite{Bilic:2001cg,Makler:2003iw,Zhu:2004aq}. On the other hand, one of the interesting applications of the polytropic structure is the accretion of matter by a charged BH inside certain gaseous fluids \cite{Jia:2024hpw}. In this study, a number of critical points for possible accretion have been explored and some limits close to the event horizon for a Maxwell-Boltzmann gas have been discovered.

In the current study, the polytropic structure model memecis, like dark energy, was described by the non-linear EoS in the form $p= -\xi \rho^{(1 + n)}$. Various cosmological questions focus on the unified dark fluid, a hybrid of dark matter and dark energy, as a candidate for the Chaplygin gas. Starting from this insight, a static, spherically symmetric Anti de-Sitter BH solution surrounded by a modified Chaplygin gas (MCG) was derived. Taking into account the equation of state of the MCG, $p = A\rho - \frac{B}{\rho \beta}$, its energy density was derived according to the radial coordinate \cite{Sekhmani:2023plr}.

In this paper, we explore a wide range of physical aspects such as QNMs, the stringent limit on grey-body factors, shadow behaviour, and the rarity of a BH surrounded by a polytropic structure that closely emulates the dark energy model. To this end, the study is structured in the following stages: The next Sec.~\ref{Model1}, together with Sec.~\ref{Model2}, is devoted to the survey of a BH physical solution surrounded by polytropic structure. In doing so, we will be able to verify a number of constraints imposed by curvature singularities and energy conditions.  In Sec.~\ref{SPP}, the sparsity of Hawking radiation is revealed. In Sec.~\ref{QNMs} we analyse GWs taking into account a massless scalar perturbation based on the Wentzel-Kramers-Brillouin (WKB) approximation method with 3rd order. The primary focus of Sec.~\ref{GF} is the investigation of grey body bonds. To proceed with the study of optical properties in terms of BH shadow analysis, Sec.~\ref{Opt} involves a complete study. In Sec.~\ref{EM}, the energy emission rate process is carried out within the polytropic parameter space. The findings and our conclusions are reported in section~\ref{Con}.
\section{Polytropic Gas Model of Dark Energy with EoS: $p =-\xi \rho^{1+\frac{1}{n}}$}
\label{Model1}
The main task considered in this section is to probe the implication of the role of a scalar gas field with a polytropic EoS in the GR. For this purpose, an adequate action needs to be thoroughly studied, which can be expressed as follows
    \begin{equation}\label{action}
    \mathcal{I}=\int d^4x\sqrt{-g}\frac{\mathcal{R}}{2\alpha^2}+\mathcal{I}_{\text{Poly}}.
\end{equation}
In this study, the Ricci scalar is represented by $\mathcal{R}$, $g_{\mu\nu}$ refers to the symmetric tensor with a determinant designated by $g=\det (g_{\mu\nu})$, and $\mathcal{I}_{\text{Poly}}$ represents the contribution of the polytropic structure. Throughout the remainder of this study, we assumed that $\alpha=8\pi G=1=c$, where $G$ and $c$ are respectively the Newtonian gravitational constant and the speed of light.

It follows that the variation of the action $(\ref{action})$ with respect to the metric tensor $g_{\mu\nu}$ gives rise to the following field equations,
 \begin{align}
 \mathcal{I}_{\mu\nu}=\mathcal{R}_{\mu\nu}-\frac{1}{2}g_{\mu\nu}\mathcal{R}-\, \mathcal{T}_{\mu\nu}^{\text{Poly}}&=0,\hspace{-0.3cm}
 \end{align}
where $\mathcal{T}_{\mu\nu}^{\text{Poly}}$ is the energy-momentum tensor for the polytropic structure.

Exploring the impact of the polytropic structure within the GR leads us to consider a static, spherical and symmetrical four-dimensional metric ansatz with $g_{tt}\,g_{rr}=-1$, given by 
 \begin{align}\label{met}
    \mathrm{d}s^2=-g(r)\,\mathrm{d}t^2+g(r)^{-1}\,\mathrm{d}r^2+r^2\,\mathrm{d}\Omega^2,
 \end{align}
where $g(r)$ is a metric function to be determined and $\mathrm{d}\Omega^2$ describes the line element of a $2$-dimensional unit sphere with a curvature constant of $2$. It can be represented by
    \begin{eqnarray}
        \mathrm{d}\Omega^2 =\mathrm{d}\theta^2+\mathrm{d}\phi^2\,\sin^2\theta.
    \end{eqnarray}
with $t\in(-\infty,+\infty)$, $r\in(0,\infty)$, $\theta\in[0,\pi]$, and $\phi\in[0,2\pi].$

It may be useful, for the sake of the present study, to envisage the perfect fluid state, which is typified by the stress-energy tensor,
\begin{equation}
    \mathcal{T}_{\mu\nu}=\left(\rho+p\right)u_\mu u_\nu+ p g_{\mu\nu}\,.
\end{equation}
In this case, $\rho$ and $p$ are respectively the energy density and the isotropic pressure, which are gauged by an observer moving with the fluid, and $u_\mu$ corresponds to its four-velocity vector. Numerous studies are currently being carried out in the field of GR, featuring spherically symmetric static solutions with the surrounding perfect fluid (dust, radiation, dark energy, or ghost energy) with an EoS $p = \omega \rho$  \cite{Kiselev:2002dx,Li:2014ixn,Setare:2007jw,Benaoum:2012uk,Bilic:2002chg,Chen2005:qnm}. There is also convincing proof that the perfect cosmological fluid surrounding the BH can be regarded as an anisotropic fluid due to the gravitational effect. It ensures that the polytropic structure has to emerge as an anisotropic fluid from various points of view, one of which claims a scalar tachyon field $\phi $ which is thought to be the source of dark energy, and a tachyon field potential $V(\phi)$, with the Born-Infeld type Dirac Lagrangian $\mathcal{L}_\phi=-V(\phi)\sqrt{1-g^{\mu\nu}\partial_\mu\nu\phi\phi}$ \cite{Garousi:2000tr}. On the other hand, following concepts based on a scalar field of $k$-essence, the polytropic structure can be rebuilt as a scalar field gas. This reconstruction is produced by looking at the action of the scalar field of $K$-essence. $S=\int \mathrm{d}^4x\sqrt{-g}\, p(\phi,\chi)$ with $p(\phi,\chi)$ being the Lagrangian density \cite{MagalhaesBatista:2009cus,Raposo:2018rjn}. Based on the above, the polytropic structure is properly modeled such that its radial pressure is different from the tangential pressure. These are anisotropic fluids, which entails a covariant form of the stress-energy tensor for the polytropic structure, as follows:
\begin{equation}\label{T1}
    \mathcal{T}_{\mu\nu}=\left(\rho+p_t\right)u_\mu u_\nu-p_t g_{\mu\nu}+\left(p_r-p_t\right)\chi_\mu \chi_\nu\,,
\end{equation}
in which the radial pressure in the direction of $\chi_\mu$ means $p_r$, while the tangential pressure orthogonal to $\chi_\mu$ is none other than $p_t$ and $\chi_\mu$ is the unit space vector orthogonal to the velocity $u_\mu$. It further turns out that $u_\mu$ and $\chi_\mu$ should satisfy the condition $u_\mu u^\mu=-\chi_\mu \chi^\mu=1$. 

We consider the frame to be co-moving with the fluid, such that $u^a=\sqrt{g(r)}\,\delta_0^a$ and $\chi^a=1/\sqrt{g(r)}\,\delta_1^a$. Starting from this premise, the stress-energy tensor \eqref{T1} takes on the following form,
\begin{equation}\label{T2}
    \mathcal{T}_\mu^\nu=-\left(\rho+p_t\right)\delta_\mu^0 \delta^\nu_0+p_t\delta_\mu^\nu+\left(p_r-p_t\right)\delta_\mu^1\delta_1^\nu\,.
\end{equation}
Here, the term $p_r-p_t$ is designated as the anisotropic factor from which its disappearance allows Eq. \eqref{T2} to describe a standard isotropic background.

To obtain a complete description of the polytropic structure as an anisotropic fluid, we consider the scalar field gas to be in a state across an event horizon defined by the stress-energy tensor \eqref{T2}. It is worth pointing out that inside the horizon where $g_{rr} <0$ and $g_{tt} >0$, the behavior of the spatial component $r$ is the same as that of the temporal component $t$. The energy density is thus $\mathcal{T}_r^r= p_r$, while the pressure in the spatial direction $t$ is expressed by $\mathcal{T}_t^t=-\rho$. From this correspondence, the energy density and pressure remain continuous provided that the requirement $p_r=-\rho$ is satisfied. In the case of $p_r\neq-\rho$ and $\rho(r_h)\neq 0$, though, the pressure at the horizon is discontinuous and the phase of the solution varies dynamically.

In the following, we will only cover the situation where $p_r=-\rho$ \cite{Kiselev:2002dx}, so that the polytropic structure is static, and, under restrictions on the solution, the energy density is continuous across the horizon. Adopting the ideas of anisotropic fluids, the tangential pressure $p_t$ is constrained by taking the isotropic mean over the angles and stating that $\braket{\mathcal{T}_i^{j}}=p(r) \delta_i^j$. In this way we can obtain
\begin{equation}\label{ptt}
   p(r) =p_t+\frac{1}{3}\left(p_r-p_t\right),
\end{equation}
where the formula $\braket{\delta_i^1 \delta_1^j}\equiv\frac{1}{3}$ is used. Conceptually, the standard tangential pressure expression for quintessential dark energy can be described using the standard formulation of Eq. (\ref{ptt}) such that $p_t =\frac{1}{2}\left(3\omega+1\right)\rho$, which is per the radial pressure $p_r = -\rho$.

The polytropic structure is characterized by a non-linear EoS $p = -\xi \rho^{1+\frac{1}{n}}$, with $\xi$ a positive parameter. Using the equation $p_r = -\rho$, the tangential pressure of the polytropic structure may be calculated as $p_t =\frac{1}{2}\rho(r)-\frac{3}{2}\xi \rho^{1+\frac{1}{n}}$. As a result, we may express the components of the polytropic structure's stress-energy tensor as follows:
\begin{align} 
    \mathcal{T}_t^t&=\mathcal{T}_r^r=-\rho\label{m1},\\
\mathcal{T}_{\theta_{1}}^{\theta_{1}}&=\mathcal{T}_{\theta_{i}}^{\theta_{i}}=\frac{1}{2}\rho(r)-\frac{3}{2}\xi \rho^{1+\frac{1}{n}}\label{m2}.
\end{align}
We further demonstrate that the anisotropy of the polytropic structure fades and that the EoS $p =-\xi \rho^{1+\frac{1}{n}}$ is maintained on the cosmological scale.
\section{Exact analytical solutions}
\label{Model2}
Since the spacetime is spherically symmetric and static, the requirement $\mathcal{T}_t^t =\mathcal{T}_r^r$ has to be fulfilled. As a result, the gravitational field equations are explicitly given as

    \begin{eqnarray}
\mathcal{I}_t^t=\mathcal{I}_r^r&=&\frac{g'(r)}{r}+\frac{1}{r^2}(g(r)-1),\label{g1}\\
     \mathcal{I}_{\theta}^{\theta}= \mathcal{I}_{\phi}^{\phi}&=&\frac{g''(r)}{2}+\frac{g'(r)}{r}.\label{g2}
    \end{eqnarray}

To probe the surrounding polytropic solution, it ought to consider the gravitational field equations (\ref{m1}-\ref{m2}) with the polytropic fluid fields equations (\ref{g1}-\ref{g2}). Thus, quite a few computations based on the conservation law related to the polytropic structure provide the following first-order differential equation:
\begin{equation}\label{ut}
    r\,\rho'(r) +3\,\rho(r)-3\,\xi\,\rho(r)^{1+\frac{1}{n}} =0\,,
\end{equation}
where the prime represents a first derivative with respect to the radial variable $r$. Thus, the energy density of the polytropic structure is solved precisely by the equation \eqref{ut} which is explicitly stated as follows
\begin{equation}
   \rho(r) =\left(A^2\, r^{\frac{3}{n}}+\xi\right)^{-n} .\label{ho}
\end{equation}
In this context, $A$ is a normalisation factor holding information about the intensity of the polytropic fluid matter. Upon further scrutiny, equation (\ref{ho}) implies the conservation law outcome of the stress-energy tensor $\partial_\mu T^{\mu\nu}=0$. 

Note that at certain limits, the polytropic energy density is altered. At large radial coordinates (i.e. $A^2 r^{\frac{3}{n}}\gg \xi$), we obtain
\begin{equation}
     \rho(r)\sim  A^{-2n}/r^3\,,
\end{equation}
which infers that the polytropic structure looks like a positive cosmological constant in a large-scale setting and that the nearer it is to the BH, the greater its gravitational clumping. For small radial coordinates (i.e., $A^2 r^{\frac{3}{n}}\ll \xi$), we are able to obtain 
\begin{equation}
     \rho(r)\sim 1/ \xi^n  \,,
\end{equation}
indicating that the polytropic scalar field gas appears as a content of matter with an energy density that varies with $r^{3}$. 
\begin{widetext}
\begin{center}
\begin{table}[t]
    \centering
    \caption{Characteristics of quintessence, Chaplygin dark fluid (CDF) and polytropic structures.}
     \scalebox{0.9}{   \begin{tabular}{@{}c||c|c|c|c|c@{}}
         anisotropic fluid  &  EoS  & $p_r$ & $p_t$ & $\rho$ & asymptotic behavior \\
         \hline\hline
       \text{Quintessence} \,&\, $p=\rho \omega$ $\left(-1<\omega<-1/3\right)$&\, $-\rho$ \,&\, $\frac{1}{2}\left(3\omega+1\right)\rho$ \,&\, $\frac{6c\,\omega}{4r^{3(\omega+1)}}$ \,&\, $\begin{array}{lcl}
 \rho&\rightarrow&0\\
 p_r&\rightarrow&0\\
 p_t&\rightarrow&0
 \end{array}$   \\
 \hline
       \text{CDF} \,&\, $p=-\gamma/\rho$ $\left(\gamma>0\right)$ \,&\,$-\rho$ \,&\, $\frac{1}{2}\rho-\frac{3\gamma}{2\rho}$ \,&\, $\sqrt{\gamma+\frac{Q^2}{r^{6}}}$ \,&\, $\begin{array}{lcl}
 \rho&\rightarrow&\sqrt{\gamma}\\
 p_r&\rightarrow&-\sqrt{\gamma}\\
 p_t&\rightarrow&-\sqrt{\gamma}
 \end{array}$   \\
       \hline
       \text{Polytropic} \,&\, $p=-\xi \rho^{1+\frac{1}{n}}$ $\left(\xi>0\right)$ \,&\,$-\rho$ \,&\, $\frac{1}{2}\rho-\frac{3}{2\xi\rho^{1+\frac{1}{n}}}$ \,&\, $\left(A^2\, r^{\frac{3}{n}}+\xi\right)^{-n}$ \,&\, 
       $\begin{array}{lcl}
 \rho&\rightarrow&\frac{A^{-2n}}{r^{3}}\\
 p_r&\rightarrow&-\frac{\xi^{-2n}}{r^{3}}\\
 p_t&\rightarrow&-p_r\bigg(3^{-1}+2\xi p_r^{1/n}\bigg)
 \end{array}$ 
          \\
    \end{tabular}}
    
    \label{de}
\end{table}
\end{center}
\end{widetext}

Choosing $r\rightarrow \infty$, on the other hand, turns out to give $p_r\rightarrow-\xi^{-n}$ and $p_{\theta,\phi} \rightarrow-\xi^{-n}$, showing that the polytropic structure is isotropic and fulfills $p=-\xi/\rho^{1+\frac{1}{n}}$ exclusively on the cosmological scale. It is noteworthy that a cosmological fluid obeying a generic EoS such that $p(\rho)=-\xi/\rho^{1+\frac{1}{n}}$, in which the radial pressure is sufficient to satisfy the constraint $p_r = -\rho$ when it encloses a central BH, thus, showing isotropic tendencies on the cosmological scale. Based on Tab. \ref{de}, we work out the large-distance limits of the pressure components in the polytropic structure, proving that the anisotropic factor cannot asymptotically reach zero, unlike the effects found in the quintessential fluid and the CDF, where the anisotropy declines with distance. In the case of the polytropic structure, the anisotropy factor falls to zero at $\xi=4(r^{3/n}A^2)/3$. This ongoing anisotropy leads to a series of implications for the evolution of the universe, notably the Hubble expansion anisotropies and the mass mechanism fluxes. These insights are underpinned by a whole series of observations, such as those targeting clusters of galaxies and type Ia supernovae (SN Ia) in the context of the Lemaitre-Tolman-Bondi (LTB) models. 
      
We now seek the BH solution surrounded by polytropic structure from the considerations of the field equations~(\ref{m1}-\ref{m2}) and~(\ref{g1}-\ref{g2}) and the resultant expression for the polytropic energy density (\ref{ho}). So, the field equation $(t,t)$ provides the differential equation:
    \begin{eqnarray}
    \nonumber
      \bigg(r\, g'(r)+g(r)-1\bigg)+r^2 \bigg(A^2\, r^{\frac{3}{n}}+\xi\bigg)^{-n}=0\,, 
    \end{eqnarray}
which leads through the procedure of solving the first-order differential equations to an analytical solution for $g(r)$ in the following form
\begin{figure*}[htb!]
      	\centering{
      	\includegraphics[scale=0.9]{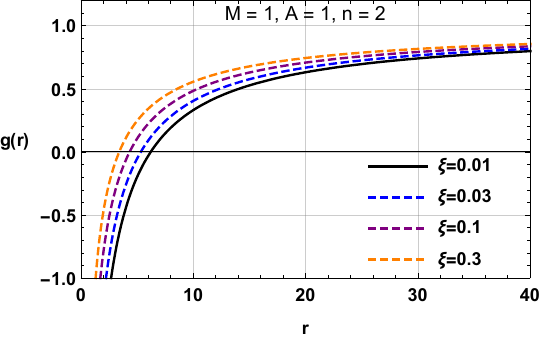}
       \hspace{5mm}
            \includegraphics[scale=0.9]{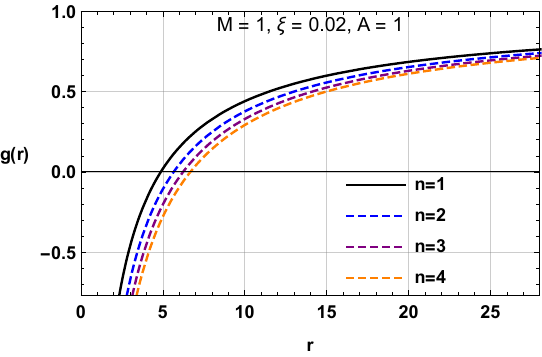}\\
            \includegraphics[scale=0.9]{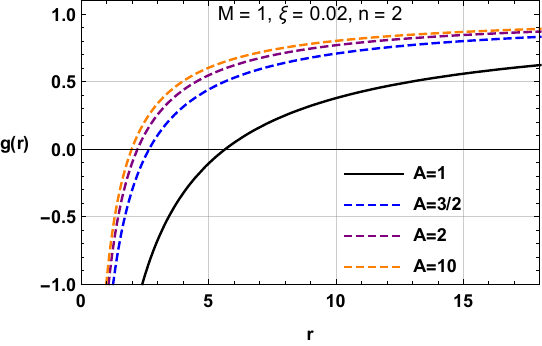}
            \hspace{5mm}
             \includegraphics[scale=0.9]{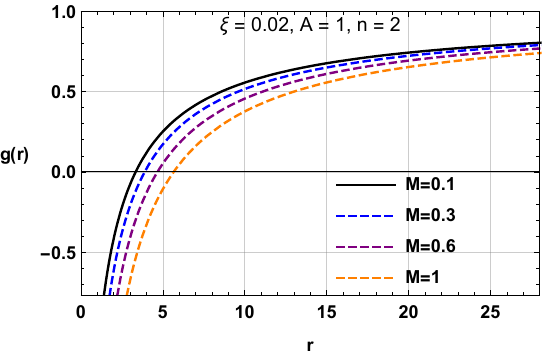}
            }
      	\caption{The metric function $g(r)$ representation \eqref{metric} as a function of $r$ for various value of the  parameter space.}
      	\label{ff1}
      \end{figure*}
    \begin{equation}\label{metric}
 g(r) =1-\frac{2M}{r}-\frac{1}{3} r^2 \xi ^{-n} \, _2F_1\left(n,n;n+1;-\frac{A^2 r^{3/n}}{\xi }\right)\hspace{0.6cm}
\end{equation}
where $M$ is the relevant physical mass for the BH solution. In this BH solution, the hypergeometric function ${_2}F_1[\alpha, \nu; \lambda; \xi]$ represents the regular solution of the hypergeometric differential equation, which is set for $\lvert \xi\rvert<1$  through a power series of the form
\begin{equation}
{_2}F_1[\lambda_1, \lambda_2; \lambda_3; \lambda_4]=\sum_{k=0}^{\infty}\bigg[(\lambda_1)_k(\lambda_2)_k/(\lambda_3)_k\bigg]\lambda_4^k/k!
\end{equation}
with $(n)_k$ is the (rising) Pochhammer symbol~\cite{Blumlein:2021hbq}.

Exploring the asymptotic behavior of the metric function $g(r)$ requires us to assume the limit $r\rightarrow\infty$. We then have
\begin{equation}
    g(r)\rightarrow 1.
\end{equation}
More detailed analysis revealed that the BH solution can be reduced to the Schwarzschild BH along with the consideration of the limit $\xi\rightarrow\infty$ such that
\begin{equation}
    g(r)\mid_{\xi\rightarrow\infty}=1-\frac{2M}{r}.
\end{equation}

\begin{figure*}[htb!]
      	\centering{
      	\includegraphics[scale=0.9]{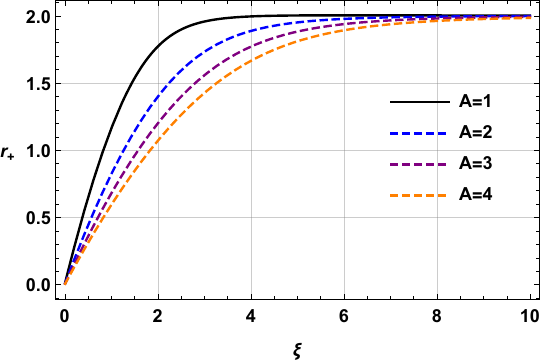}
       \hspace{5mm}
            \includegraphics[scale=0.9]{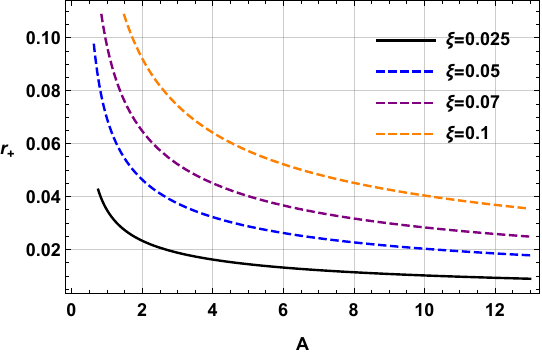}\\
            \includegraphics[scale=0.75]{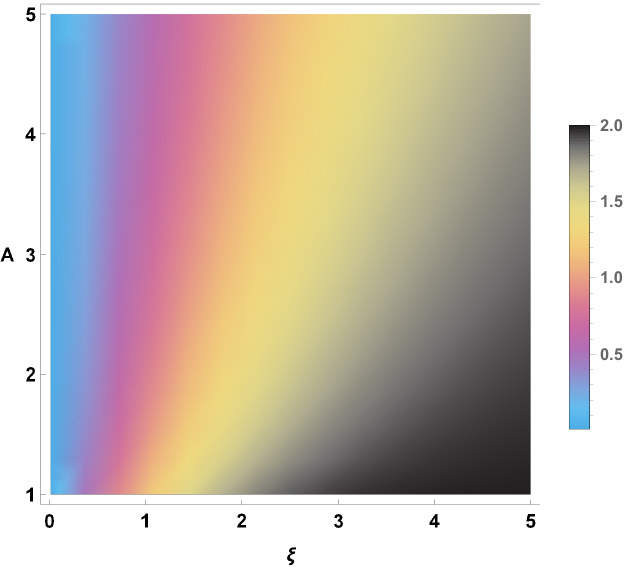}
            }
      	\caption{The horizon radius behaviors as a function of the pair polytropic parameter $(\xi,A)$ for $n=3$ (upper row). The extremal BH mass with varying $A$ and $\xi$ for $n=3$ (lower row).}
      	\label{ff2}
      \end{figure*}

For a more in-depth analysis of the behaviour of the BH in relation to the parameter space of the polytropic structure, Fig. \ref{ff1} provides an appropriate observation in terms of the $(g(r), r)$ plane. Thus, it can be observed that for each parameter variation on the corresponding parameter space of the BH $(\xi,A,n,M)$, the set of possible horizon radii is constrained to a single real root, which is nothing other than the event horizon radius $r_+$. On the other hand, using the mass $M$ as the unit, the event horizon radius $r_h$ in relation to pair parameters of the polytropic structure is depicted in Fig. \ref{ff2} for different values of each parameter, namely, $\xi$ and $A$. In addition, We plot the extremal BH mass with varying $A$ and $\xi$

Subsequently, we will employ tools to forecast the curvature singularities to examine the properties of the BH solution. Furthermore, we need an analysis made up of scalar invariants to provide an appropriate uniqueness and singularity proof of our BH solution. These types of information are, respectively, the Ricci scalar, the squared Ricci, and the Kretschmann scalar, which are provided by

\begin{align}
    \mathcal{R}&=\frac{\xi ^{-n} (\frac{A^2 r^{3/n}}{\xi }+1)^{-n} \left(A^2 r^{3/n}+4 \xi \right)}{A^2 r^{3/n}+\xi },\label{R}
\\
\mathcal{R}_{\alpha\beta}\mathcal{R}^{\alpha\beta}&=   \frac{\xi ^{-2 n} (\frac{A^2 r^{3/n}}{\xi }+1)^{-2 n} \left(4 A^2 \xi  r^{3/n}+5 A^4 r^{6/n}+8 \xi ^2\right)}{2 \left(A^2 r^{3/n}+\xi \right)^2},\label{RR}
\\
\mathcal{R}_{\alpha\beta\mu\nu}\mathcal{R}^{\alpha\beta\mu\nu} &=\frac{4}{3} \xi ^{-2 n} \, _2F_1\left(n,n;n+1;-\frac{r^{3/n} A ^2}{\xi }\right) \bigg(\frac{12 M \xi ^n}{r^3}-\frac{(\frac{A ^2 r^{3/n}}{\xi }+1)^{-n} \left(5 A ^2 r^{3/n}+2 \xi \right)}{A ^2 r^{3/n}+\xi }\bigg),\nonumber\\
&+\frac{4}{3} \xi ^{-2 n}
   \, _2F_1\left(n,n;n+1;-\frac{r^{3/n} A ^2}{\xi }\right)^2+\frac{48 M^2}{r^6}-\frac{8 M \xi ^{-n} (\frac{A ^2 r^{3/n}}{\xi }+1)^{-n} \left(5 A ^2 r^{3/n}+2 \xi \right)}{r^3 \left(A ^2 r^{3/n}+\xi \right)}\nonumber\\
   &+\xi ^{-2 n} \bigg(\frac{A
   ^2 r^{3/n}}{\xi }+1\bigg)^{-2 n} \bigg(\frac{9 A^4 r^{6/n}}{\left(A ^2 r^{3/n}+\xi \right)^2}+4\bigg).\label{RRR}\end{align}

Looking into the terms \eqref{R}, \eqref{RR}, and \eqref{RRR} points out that the BH solution described by the line element is strictly singular for each valued parameter space. The singularity can occur for the metric function solution only through the mass term, while the polytropic term vanishes, and then the metric function remains constant. However, in the remaining analysis, we shall adhere to the metric function \eqref{metric} and not address any further scenario. The singularity phenomenon is demonstrated using the Ricci scalar, squared Ricci, and Kretschmann scalar at the center $r=0$ to give an in-depth illustration. The subsequent findings are obtained:
\begin{align}
    \lim\limits_{r\to 0} \mathcal{R} &\approx 4\xi ^{-2 n}\,,\\
    \lim\limits_{r\to 0}\mathcal{R}_{\alpha\beta}\mathcal{R}^{\alpha\beta} &\approx\infty\,\, \left(\textit{if}\,;\,\, n<\frac{3}{4}\right)\,,\\
    \lim\limits_{r\to 0}\mathcal{R}_{\alpha\beta\mu\nu}\mathcal{R}^{\alpha\beta\mu\nu} &\approx\infty\,,
\end{align}

Furthermore, investigating long-distance behavior offers a useful additional perspective, given that
\begin{align}
    \lim\limits_{r\to \infty} \mathcal{R} &\approx 0\,,\\
     \lim\limits_{r\to \infty}\mathcal{R}_{\alpha\beta}\mathcal{R}^{\alpha\beta} &\approx0\,,\\
    \lim\limits_{r\to \infty}\mathcal{R}_{\alpha\beta\mu\nu}\mathcal{R}^{\alpha\beta\mu\nu} &\approx0,
\end{align}
which infers that the Ricci scalar, the Ricci squared, and the Kretschmann scalar have a finite term as their long-range. In conclusion, the scalar tools demonstrate that the BH solution we have obtained is unique, and the polytropic background significantly modifies the BH spacetime.
\subsection{Energy Conditions}
To shed some light on the behavior of our BH solution, we now turn to analyze the classical energy conditions (ECs), namely the null energy condition (NEC), the dominant energy condition (DEC), the weak energy condition (WEC), and the strong energy condition (SEC), which are given as follows~\cite{Kontou:2020bta}:
\begin{figure*}[!htp]
      	\centering{
       \includegraphics[scale=0.9]{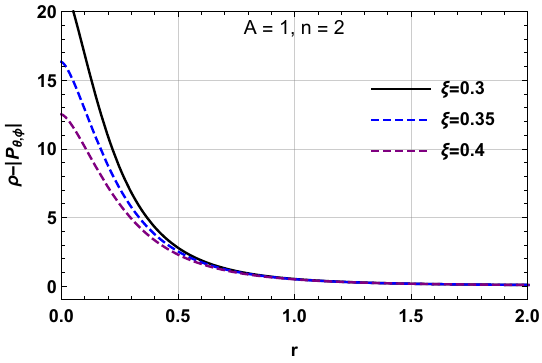} \hspace{2mm}
      	\includegraphics[scale=0.9]{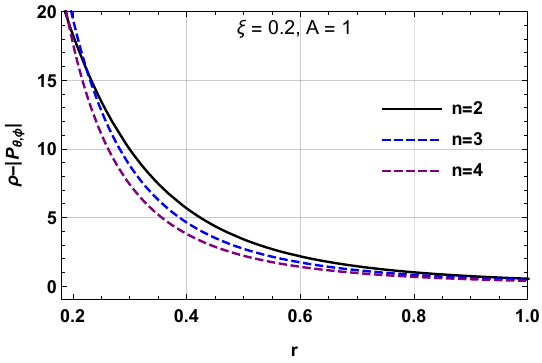} \hspace{2mm}
      	}\\
      	\centering{
       \includegraphics[scale=0.9]{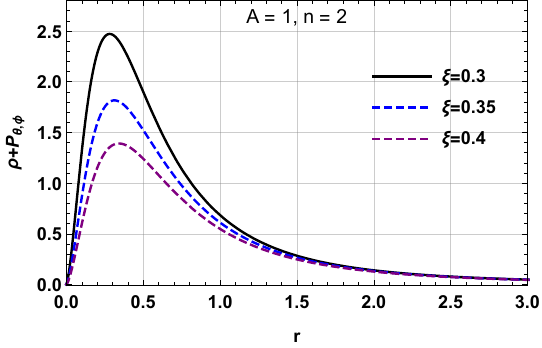} \hspace{2mm}
      	\includegraphics[scale=0.9]{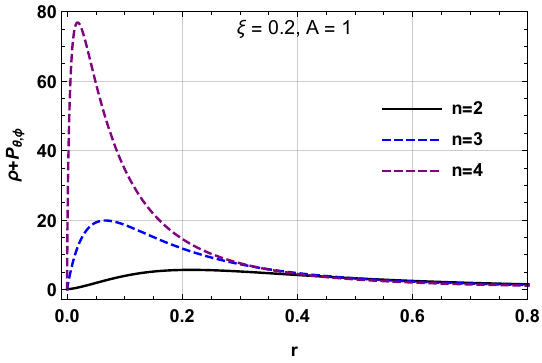} \hspace{2mm}
      }
      	\centering{
       \includegraphics[scale=0.9]{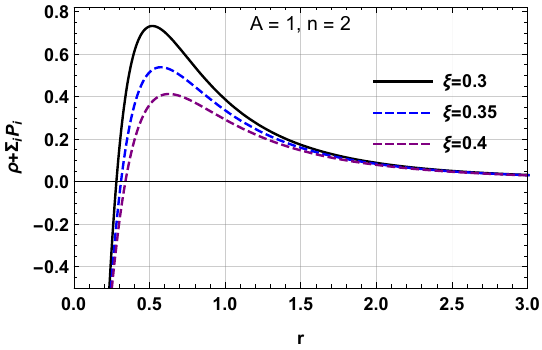} \hspace{2mm}
       \includegraphics[scale=0.9]{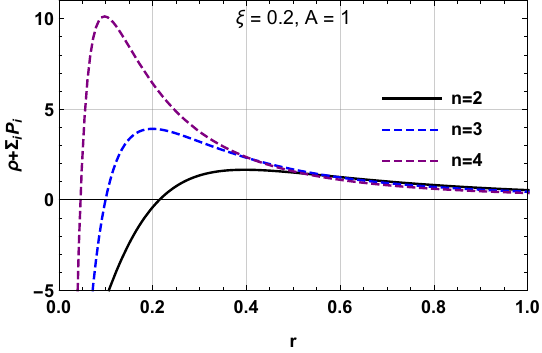} \hspace{2mm}
      	}
      	\caption{The variation of$\rho+\sum_i p_i$ (strong energy condition), $\rho+p_{\theta,\phi} $ (null energy condition), and $\rho-\mid p_{\theta,\phi}\mid $ (dominant energy condition) against $r$ for various value of the parameters $\xi$ and $n$.}
      	\label{ff3}
      \end{figure*}

\begin{eqnarray}\label{18}
\textbf{WEC}&:& \rho\geq 0,\; \rho+p_i\geq 0,\nonumber\qquad
\textbf{SEC}: \rho+\sum_{i}p_i\geq 0,\, \;\rho+p_i\geq 0,\nonumber \\[2mm]
\textbf{NEC}&:& \rho+p_i\geq 0,\nonumber \qquad\qquad \quad\label{19}
\textbf{DEC}: \rho\geq 0, \;|p_i|\geq \rho.
\label{20}
\end{eqnarray}
Correspondingly, the key expressions can be supplied as follows:
\begin{align}
   \rho + p_r =0,\,\, \rho+ p_{\theta,\phi}&=\frac{3}{2}\bigg(\rho-\xi \rho^{1+\frac{1}{n}}\bigg),\nonumber\\
   \rho+ \sum_{i}p_i&=\rho-\frac{3}{2}\xi \rho^{1+\frac{1}{n}}\,,\\
   \rho-|p_r|=0,\,\,  \rho-|p_{\theta,\phi}|&=\rho-\Bigl|\frac{1}{2}\rho-\frac{3}{2}\xi \rho^{1+\frac{1}{n}}\Bigr|\,.\nonumber
\end{align}

A thorough examination of the ECs satisfaction/violation clearly reveals that the SEC can be satisfied provided that the following constraint is met,
\begin{equation}\label{secc}
 \xi  \left(A ^2 r^{3/n}+\xi \right) \leq\frac{2}{3},
\end{equation}
where a violation of \eqref{secc} implies that the SEC property is not satisfied. 

A similar analysis is performed to ascertain satisfaction/violation of the DEC, revealing that the DEC is still being satisfied, considering
\begin{equation}
    0\leq\frac{1}{2} \left(\xi +A^2 r^{3/n}\right)^{-n-1} \left(4 \xi +A^2 r^{3/n}\right).
\end{equation}
Furthermore, the NEC constraints can be satisfied in the context of ECs arguments provided that the following conditions are satisfied:
\begin{equation}
\xi^{n+1}\leq   0
\end{equation}
To emphasise the satisfaction or unsatisfaction constraints related to ECs in the essence of the polytropic BH framework, Fig.~\ref{ff3} depicts the variation of $\rho+\sum_{i}p_i$, i.e., SEC, $\rho+ p_{\theta,\phi}$, i.e., NEC, and $\rho-|p_{\theta,\phi}|$, i.e., DEC against the radial spacetime variable $r$. To this end, interesting observations have shown that $\rho+ p_{\theta,\phi}$ and $\rho-|p_{\theta,\phi}|$ are positive definite quantities with respect to certain spectrum variations of the parameters $\xi$ and $n$. On the other hand, $\rho+\sum_{i}p_i$ shows a change of sign at $r_{crit}=2^{-n/3} \left(\xi / A ^2\right)^{n/3}$ which involves a transition between a negative sign for small $r$ and a positive sign for large $r$. More concretely, the root $r_{crit}$ behaves like the transition point where $p_{\theta,\phi}$ changes sign, i.e., the tangential pressure changes from a repulsive to an attractive state. Overall, the assessment and evaluation revealed that the polytropic structure satisfies the NEC, WEC, and DEC criteria, whilst not satisfying the SEC. From a cosmological point of view, it has been claimed that the violation of SEC in GR is none other than the unsatisfying behaviour of gravity. Nonetheless, this approach is unlikely to have generalised validity in extended gravity, as demonstrated in~\cite{Santos:2016vjg} by looking at the $f(R)$ model.

To have an overview of what is involved in the violation or satisfaction of a certain type of fluid matter in the context of GR, it is worth noting to consider some works published in the literature. For this reason, logotropic fluids in the context of anti-de Sitter (AdS) BHs have been shown to violate the SEC for sufficiently large radii~\cite{Capozziello:2022ygp}. Similarly, the regular Hayward-AdS spacetime was found to satisfy the WEC while violating the SEC~\cite{Fan:2016rih}. In addition, some newly invented solutions for regular BHs supplied with multi-horizons have been suggested in~\cite{Rodrigues:2020pem}. It states that the SEC cannot be satisfied within the event horizon regardless of solutions, whereas the other ECs rely on the ratio between isolated solutions' extreme charges. For the sake of both rotating and non-rotating BHs in conformal gravity, an analogous finding is suggested in~\cite{Toshmatov:2017kmw}, in which the SEC can only be satisfied for particular BH sizes that rely on the intrinsic scale of the model.


\section{Sparsity of Hawking radiation}\label{SPP}
In this section, we aim to study the sparsity of Hawking radiation in the essence of our BH solution. Essentially, a BH can act similarly to a black body, emitting particles at a temperature close to the surface gravity. Even so, the Hawking radiation flux is quite dissimilar to conventional blackbody radiation in that it emerges rather sparsely throughout the evaporation process. Sparity refers to the average time between the emission of successive quanta on time scales governed by the energy of the quanta. It can therefore be defined as follows \cite{Page:1976df,Gray:2015pma,Sekhmani:2024dyh}
\begin{equation}
\label{defSpars}
    \eta =\frac{\mathcal{C}}{\Tilde{g} }\left(\frac{\lambda_t^2}{\mathcal{A}_{eff}}\right),
\end{equation}
where $\mathcal{C}$ is a dimensionless constant, $\Tilde{g}$ is the degeneracy factor of the spin of the emitted quanta, $\lambda_t=2\pi/T_H$ denotes the thermal wavelength, and $\mathcal{A}_{eff}=27 \mathcal{A}_{BH}/4$ is the related effective area of the BH. In the ordinary case of a Schwarzschild BH and the emission of massless spin-1 bosons, we have $\lambda_t=8\pi r_h^2\,\Longrightarrow\,\eta_{Sch}=64\pi^3/27\approx73.49$. By way of comparison, look at $\eta\ll 1$ for blackbody radiation.

It turns out to be advantageous, at first glance, to implement the so-called surface gravity to determine the corresponding temperature, which is expressed as follows
\begin{equation}\label{surface}
    \kappa=\left(-\frac{1}{2}\nabla_\mu\chi_\nu\nabla^\mu\chi^\nu\right)^{1/2}\bigg\vert_{r=r_+}=\frac{1}{2}g'\left(r\right)\bigg\vert_{r=r_+}
\end{equation}
where $\chi^\mu=\partial/\partial t$ is a Killing vector. Thus, to derive the pertinent surface gravity, it is convenient to look at the metric function \eqref{metric} and use the mass term which is given by
\begin{equation}
     M=\frac{ r_+}{2}-\frac{r_+^3}{6} \xi ^{-n} \, _2F_1\left(n,n;n+1;-\frac{A^2 r_+^{3/n}}{\xi }\right)
\end{equation}
into Eq. \eqref{surface}. So the surface gravity related to the BH can be accurately stated as follows
\begin{equation}
    \kappa=-\frac{r_+}{2}\left(A^2 r_+^{3/n}+\xi\right)^{-n}+\frac{1}{2r_+}.
\end{equation}
To find the corresponding Hawking temperature, we follow the formula $T=\kappa/2\pi$, which yields the temperature in such a way as
\begin{equation}\label{TTT}
    T_H=\frac{1}{8\pi r_+}\left(1-\frac{r_+^2}{2}\left(A^2 r_+^{3/n}+\xi\right)^{-n}\right).
\end{equation}
In accordance with the first law of BH thermodynamics, entropy is calculated as follows
\begin{equation}\label{r33}
    S=T^{-1}\int\frac{\partial M}{\partial r_+}\,\mathrm{d}r_+=\pi\, r_+^2.
\end{equation}

To highlight the real nature of the physical phenomenon of sparsity in BHs surrounded by a polytropic structure, we consider the relevant Hawking temperature (\ref{TTT}) and, in accordance with the definition of sparsity (\ref{defSpars}), we obtain
 \begin{figure*}[t!]
      	\centering{
      	\includegraphics[scale=0.9]{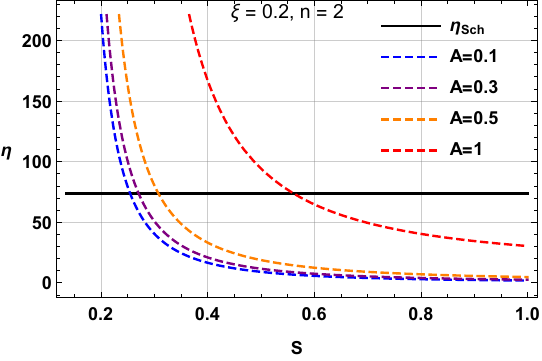} \hspace{5mm}
       \includegraphics[scale=0.9]{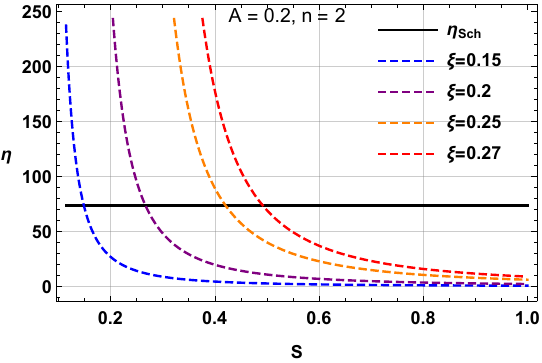}\\
       \includegraphics[scale=0.9]{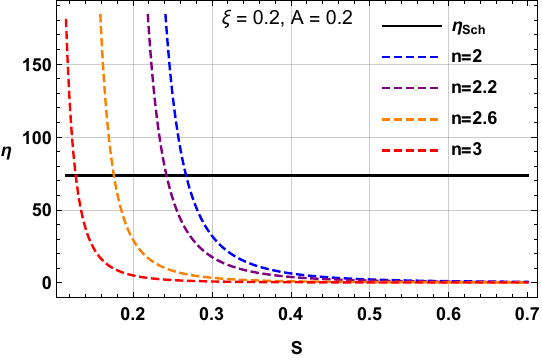}
       
       }
      	\caption{Sparsity behavior $\eta$ versus the entropy $S$ for various values of the parameter space.}
      	\label{ff5}
      \end{figure*}

\begin{equation}
    \eta=\frac{64 \pi  S^2}{27 \Bigg(\frac{S^2}{\pi^2} \bigg(\left(A ^2 \pi ^{\left.-\frac{3}{2}\right/n} S^{\left.\frac{3}{2}\right/n}+\xi
   \right)^{-n}\bigg)- \frac{S}{\pi} \Bigg)^2 },
\end{equation}
where, as expected, the full BH parameter space should affect the behaviour of the BH density. In fact, the choice of the partial set $(\xi=0,\, A=0)$ reduces the corresponding sparsity to that of the Schwarzschild spacetime.

For a better approach on how the sparsity behaviour is behaving in view of the parameter variation of the BH system, Fig. \ref{ff5} depicts suitable variation for the physical function $\eta(S)$. To begin with, the sparsity behaviour is shown for the variation of the plytropic parameters such that $\xi$, $A$, and $n$. Unlike the Schwarzschild BHs, the corresponding sparsity raises with growing in the space-$S$. As observed, at smaller $S$, the corresponding sparsity is higher than the standard sparsity of the Schwarzschild BHs, which provides the interpretation that at the evaporation phase, the emitted radiation is sparser than the Hawking radiation. By pursuing this process, for $S$ large enough, the corresponding $\eta$ decreases monotonically and approaches zero through an asymptotic way. Thus, according to this phase, the corresponding behaviour of the sparsity becomes more identical to the black-body radiation. It turns out, rather interestingly, that the polytropic parameter set such that $\xi$, $A$, and $n$ assigns the decay rate in a non-trivial way. A closer comparative study reveals that as the parameters $\xi$ and $A$ decrease, the corresponding sparsity appears to shrink more sharply for large $S$. On the other hand, varying the plytropic index $n$ as a physical parameter in an increasing sense causes a more immediate shrinkage of the corresponding sparsity.


\section{Ring-Down GWs or Quasinormal Modes}\label{QNMs}
In this section, we examine the behavior of massless scalar perturbations in the spacetime of a BH with a surrounding polytropic structure, presuming that the test field has a trivial impact on the spacetime of the BH. This simplification enables us to focus on the characteristics associated with perturbations. To analyze these perturbations, we obtain the relevant Schr\"odinger-like wave equations that transform into Klein-Gordon-type equations in the context of massless scalar fields. This transformation makes it possible to adopt relations that are relevant to the spacetime under consideration, thereby guaranteeing a consistent mathematical context. The principal purpose of our study is to explore QNMs, which reflect the characteristic vibrations of a BH during perturbation. To this end, we apply the WKB approach up to third order. Using this approach is particularly appropriate for the treatment of wave equations in the curved, perturbed spacetime surrounding a BH.In addition, this approach implies a thorough investigation of the potential barrier generated by the perturbations near BH and the behavior of the scalar field therein. This helps us to precisely work out the frequencies and damping rates of QNMs, yielding vital insight into the stability and dynamic resilience of BHs to exterior perturbations.

The perturbed metric in accordance with the axial perturbation is given by \cite{Bouhmadi-Lopez:2020oia, Gogoi:2023kjt}:
\begin{equation}
    \label{pert_metric}
ds^2 =  r^2 \sin^2\!\theta\, (d\phi - p_2(t,r,\theta)\, dr -  p_1(t,r,\theta)\,
dt  - p_3(t,r,\theta)\, d\theta)^2 + g_{rr}\, dr^2 -\, |g_{tt}|\, dt^2 +
r^2 d\theta^2,
\end{equation} 
where the involved parameters within the perturbed metric, namely $p_1$, $p_2$, and $p_3$, can identify the perturbations to the BH spacetime. In turn, the classical metric functions $g_{tt}$ and $g_{rr}$ are the unperturbed or zeroth-order terms of the BH spacetime.

At first, we assume that the massless scalar field is close to our BH solution surrounded by a polytropic background, and we take into account that the impact of the scalar field on spacetime is minimal. This consideration allows us to formulate the perturbed metric. Eq. \eqref{pert_metric} in the following form  
\begin{equation}
ds^2 = g_{rr}\, dr^2 +r^2 d \Omega^2-\,|g_{tt}|\, dt^2 .
\end{equation}
From the above consideration, we can express the Klein-Gordon equation in curved spacetime as
\begin{equation}  \label{scalar_KG}
\square \Phi = \dfrac{1}{\sqrt{-g}} \partial_\mu (\sqrt{-g} g^{\mu\nu}
\partial_\nu \Phi) = 0.
\end{equation}
Based on the above equation, we can deal with the characterization related to the massless scalar perturbation. Thus, the massless scalar field can be given by
\begin{equation}\label{scalar_field}
\Phi(t,r,\theta, \phi) = \dfrac{1}{r} \sum_{l,m} \psi_l(t,r) Y_{lm}(\theta,
\phi).
\end{equation}
where $ Y_{lm}(\theta, \phi) $ are spherical harmonics, and $ l $ and $ m $ are their indices. In the previous terms, the wave function $ \psi_l(t,r) $ refers to the time-dependent radial function. Using this decomposition \eqref{scalar_field} in the Klein-Gordon equation \eqref{scalar_KG}, we obtain:
\begin{equation}  \label{radial_scalar}
\partial^2_{r_*} \psi(r_*)_l + \omega^2 \psi(r_*)_l = V_{eff}(r) \psi(r_*)_l,
\end{equation}
where $ r_* $ is nothing more than the tortoise coordinate expressed by the following equation as 
\begin{equation}  \label{tortoise}
\dfrac{dr_*}{dr} = \sqrt{g_{rr}\, |g_{tt}^{-1}|}.
\end{equation}
The term $ V_{eff}(r) $ is the effective potential given by:
\begin{equation}  \label{Vs}
V_{eff}(r) = |g_{tt}| \left( \dfrac{l(l+1)}{r^2} +\dfrac{1}{r \sqrt{|g_{tt}|
g_{rr}}} \dfrac{d}{dr}\sqrt{|g_{tt}| g_{rr}^{-1}} \right),
\end{equation}
Here, $ l $ is the variable standing for the multipole moment concerning the QNMs of the BH.
\begin{figure*}[t!]
      	\centering{
      	\includegraphics[scale=0.9]{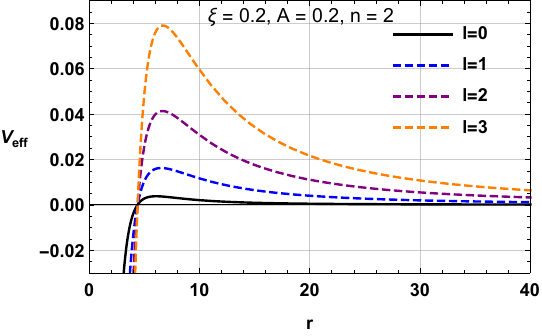}
       \hspace{5mm}
            \includegraphics[scale=0.9]{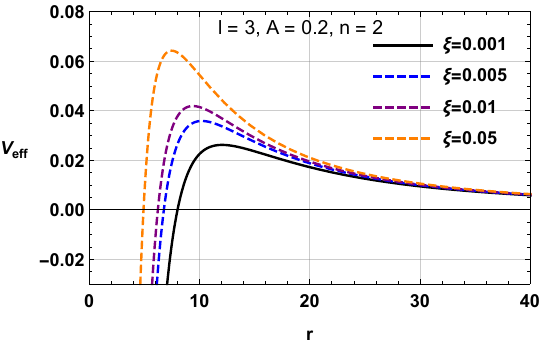}\\
            \includegraphics[scale=0.9]{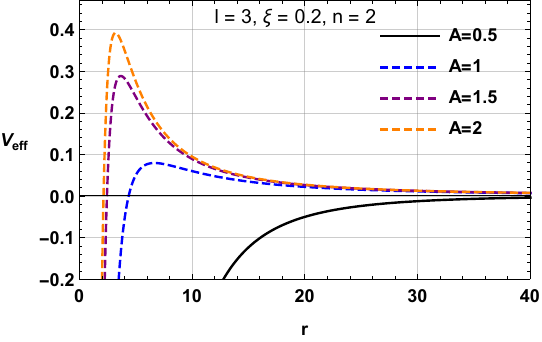}
            \hspace{5mm}
             \includegraphics[scale=0.9]{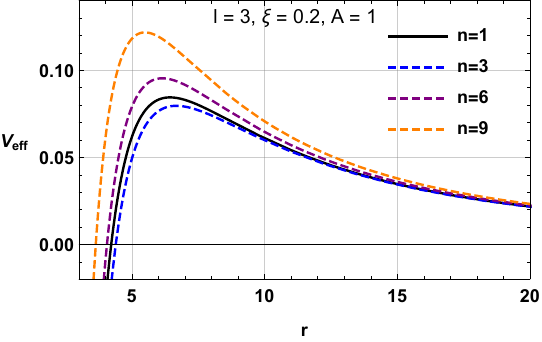
}
            }
      	\caption{Change in scalar potential as a function of $r$ using $M = 1$ for various values of the parameter space}
      	\label{ff6}
      \end{figure*}
Now we use the third-order WKB approximation method to compute the QNMs referring to our BH solution. Practically speaking, the oscillation frequency $\omega$ that characterized the GWs could be derived according to the following formula~\cite{Schutz:1985km,Iyer:1986np,Konoplya:2003ii,Matyjasek:2019eeu}:
\begin{equation}
\omega = \sqrt{-\, i \left[ (n_l + 1/2) + \bar{\Lambda}_2 + \bar{\Lambda}_3 \right] \sqrt{-2 V_0''} + V_0},
\end{equation}
where $\bar{\Lambda}_2$ and $\bar{\Lambda}_3$ are the second order and third order corrected terms, $n_l = 0, 1, 2\hdots$, stands for overtone number, $V_0 = V|_{r\, =\, r_{max}}$ and 
$V_0'' = \dfrac{d^2 V}{dr^2}|_{r\, =\, r_{max}}$. The position where the potential function $ V(r) $ attains its maximum value is labelled $ r_{\text{max}} $.


\begin{figure*}[t!]
      	\centering{
      	\includegraphics[scale=0.9]{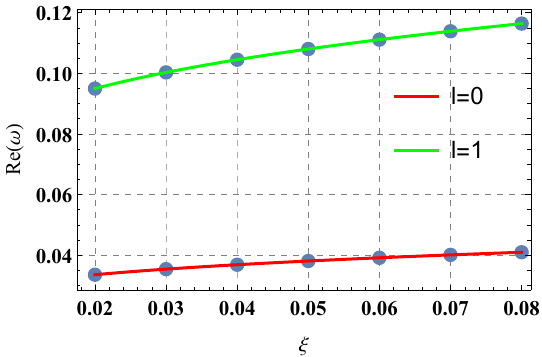}
       \hspace{5mm}
            \includegraphics[scale=0.93]{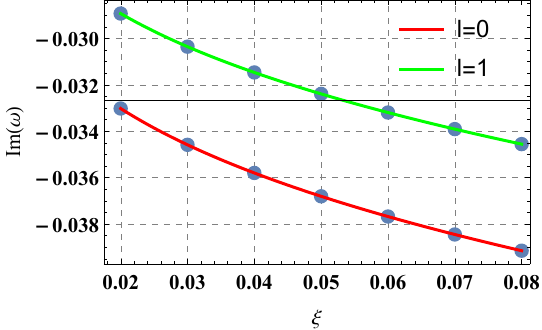}\\
            \includegraphics[scale=0.9]{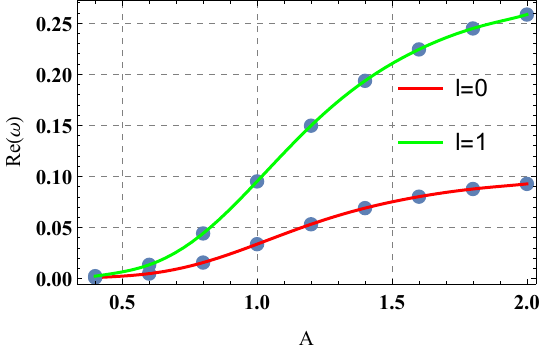}
            \hspace{5mm}
             \includegraphics[scale=0.9]{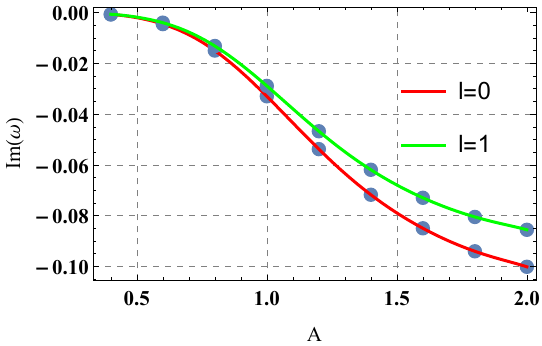}
            }
      	\caption{Variation of real and imaginary QNMs using $M = 1$ for the case where the polytropic index is fixed at $n=2$ and $n_l=0$.}
      	\label{ff7}
      \end{figure*}
\begin{figure*}[t!]
      	\centering{
      	\includegraphics[scale=0.9]{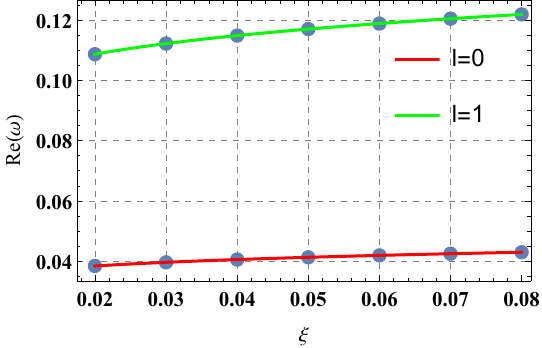}
       \hspace{5mm}
            \includegraphics[scale=0.93]{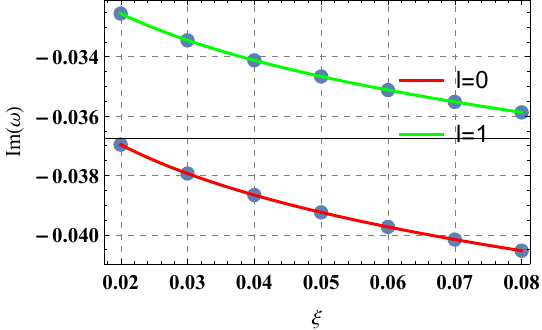}\\
            \includegraphics[scale=0.9]{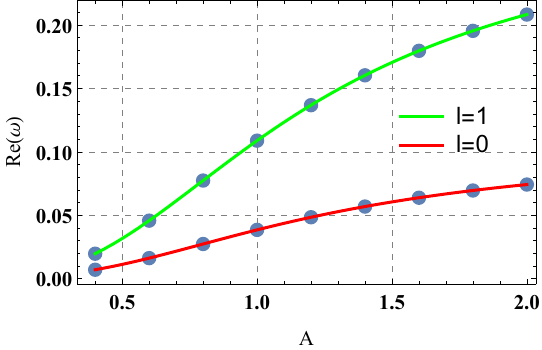}
            \hspace{5mm}
             \includegraphics[scale=0.9]{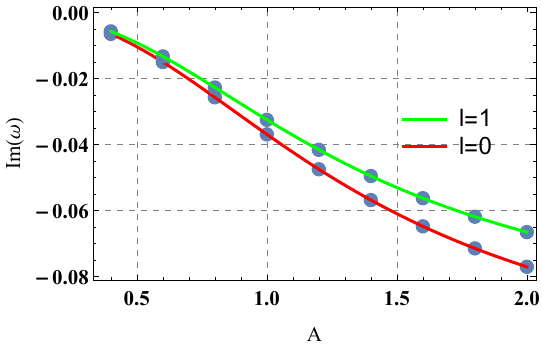}
            }
      	\caption{Variation of real and imaginary QNMs using $M = 1$ for the case where the polytropic index is fixed at $n=1$ and $n_l=0$.}
      	\label{ff8}
      \end{figure*}


To have a better understanding of how the scalar potential acts in view of the parameter variation of the BH system together with the multipole quantum number, Fig. \ref{ff6} depicts related scenarios for the function $V_{eff}(r)$. In this regard, variation in the set of the polytropic structure leads to the following results: Increasing in $\xi$ causes an increase in the potential peak and shifts it laterally towards the BH's event horizon. Similarly, it is found that also the variation of the parameter $A$ leads to increases in the peak of the potential and shifts it laterally towards the black hole's event horizon. In the same way, it is observed that the impact of the poytropic index on the variation of the scalar potential is identical to the variation of the parameters $\xi$ and $A$. Broadly speaking, these observations might be evidence that all the parameters associated with polytypic structures have a consistent effect on the spectrum of the QNMs. This is supported by explanatory computations of the QNMs modes in the following part of our study.

Tab. \ref{tut} lists the complex QNMs for different values of the overtone number $ n_l $, the multipole moment $ l $, and the pair polytropic parameter $(\xi, A)$. Notice that the discrepancies in the QNMs obtained using the first-, second-, and third-order WKB approximations diminish as $ |l - n_l| $ grows. This outline is characteristic of the WKB approximation method, which exhibits certain weaknesses in producing precise results when the overtone number $ n_l $ is higher than the multipole moment $ l $ \cite{Gogoi:2023kjt,Konoplya:2003ii}.
\begin{table}[ht!]
		\caption{QNMs for the case, $M=1$ and $n=2$.}
		\centering
		\label{tut}
		\begin{tabular}{|c|c|c|c|c|}
		\hline
		\multicolumn{5}{|c|}{A=1}\\
		\hline
		 $\xi=0.5$ & $n_l=0$ & $n_l=1$ & $n_l=2$ & $n_l=3$\\
		 \hline
		 $l=1$ & $0.169025\, -0.047794 i$ & $0.158112\, -0.147829 i$ & $0.143205\, -0.257330 i$ & $0.130888\, -0.374900 i$\\
		 \hline
		 $l=2$ & $0.280058\, -0.047542 i$ & $0.272739\, -0.144394 i$ & $0.260000\, -0.245971 i$ & $0.244821\, -0.353650 i$\\
		\hline
		 $l=3$ & $0.391458\, -0.047471 i$ & $0.386062\, -0.143338 i$ & $0.375998\, -0.241832 i$ & $0.362590\, -0.344194 i$\\
		 \hline
		\hline
		 $\xi=0.8$ & $n_l=0$ & $n_l=1$ & $n_l=2$ & $n_l=3$\\
		 \hline
		 $l=1$ & $0.189880\, -0.053796 i$ & $0.177346\, -0.166423 i$ & $0.160035\, -0.289827 i$ & $0.145377\, -0.422308 i$\\
		 \hline
		 $l=2$ & $0.314844\, -0.053531 i$ & $0.306426\, -0.162573 i$ & $0.291707\, -0.276933 i$ & $0.274019\, -0.398180 i$\\
		\hline
		 $l=3$ & $0.440173\, -0.053456 i$ & $0.433972\, -0.161403 i$ & $0.422378\, -0.272300 i$ & $0.406867\, -0.387545 i$\\
		 \hline
		\hline
		\hline
		\multicolumn{5}{|c|}{A=2}\\
		\hline
		 $\xi=0.5$ & $n_l=0$ & $n_l=1$ & $n_l=2$ & $n_l=3$\\
		 \hline
		 $l=1$ & $0.276379\, -0.090836 i$ & $0.250393\, -0.284674 i$ & $0.217551\, -0.502813 i$ & $0.194451\, -0.740476 i$\\
		 \hline
		 $l=2$ & $0.456455\, -0.090032 i$ & $0.438447\, -0.274907 i$ & $0.407866\, -0.472544 i$ & $0.373164\, -0.686361 i$\\
		\hline
		 $l=3$ & $0.637453\, -0.089801 i$ & $0.624094\, -0.271912 i$ & $0.599441\, -0.461157 i$ & $0.567276\, -0.660841 i$\\
		 \hline
		\hline
		 $\xi=0.8$ & $n_l=0$ & $n_l=1$ & $n_l=2$ & $n_l=3$\\
		 \hline
		 $l=1$ & $0.278879\, -0.091701 i$ & $0.252615\, -0.287401 i$ & $0.219422\, -0.507665 i$ & $0.196072\, -0.747667 i$\\
		 \hline
		 $l=2$ & $0.460590\, -0.090889 i$ & $0.442390\, -0.277529 i$ & $0.411480\, -0.477067 i$ & $0.376402\, -0.692960 i$\\
		\hline
		 $l=3$ & $0.440173\, -0.053456 i$ & $0.629730\, -0.274503 i$ & $0.604812\, -0.465560 i$ & $0.572302\, -0.667166 i$\\
		 \hline
		\end{tabular}
	\end{table}

To provide a more detailed insight into the investigation of massless scalar QNMs, Figs. \ref{ff7}-\ref{ff8} show the oscillation frequency, or the so-called real part of the QNMs, and the damping ratio of ring GWs according to various fixed values of the polytropic parameters $(\xi, A, n)$ for multipole moments set as $l={0,1}$. As is evident from the behaviour of the potential, the pair of polytropic parameters such that $ \xi $ and $ A $ have quite identical impacts on the QNMs. As $n=2$ is concerned, the increases of the parameter $\xi$ lead to slow increases in the oscillation frequency of GWs. On the other hand, a closer observation shows a diminishment in the damping rate as the $\xi$ raises. Thus, similar behaviour is shown for the two-valued case of the multipole moment such that $l=0,1$ and also when the situation is referred to $n=1$ (see the upper row of Fig. \ref{ff8}). In what is concerning the variation of the parameter $A$, the oscillation frequency of GWs is fastly increasing for the set where $l=0,1$ and $n=1,2$. In contrast, the damping rate described by the imaginary part of the massless QNMs spectrum rapidly decreases with increases of $A$. This behaviour is observed for all valued cases of the multipole moment as well as for the polytropic index $n$.       


\section{Greybody bounds}\label{GF}
In this section, we focus on the bounds of greybody factors. Our analysis in the previous section, based on the WKB method for QNMs, highlights the significant influence of black hole model parameters on the QNMs spectrum. Now, we investigate scalar perturbations concerning greybody factors and examine how model parameters affect these bounds using an analytical approach. Indeed, analytical techniques for predicting rigorous limits of gray body factors were first invented by Visser \cite{Visser:1998ke} and later improved by Boonserm and Visser \cite{Boonserm:2008zg}. Further studies by Boonserm et al. \cite{Boonserm:2017qcq}, Yang et al. \cite{Yang:2022ifo}, Gray et al. \cite{Gray:2015xig}, Ngampitipan et al. \cite{Ngampitipan:2012dq}, and others \cite{Chowdhury:2020bdi,Miao:2017jtr,Liu:2021xfs,Barman:2019vst,Xu:2019krv,Boonserm:2017qcq} have gone deeper into these limits. Our work broadens this investigation by analyzing a BH solution with a polytropic scalar field gas, enhancing our insight into greybody factors in various settings.

We look more closely at the constraints on grey-body factors for black holes with a surrounding polytropic structure, restricting our attention to massless scalar perturbations. To achieve this, we examine the Klein-Gordon equation underlying the massless scalar field, as outlined in the preceding section. We turn to consider the reduced effective potential, $V_{eff}(r)$, which is given by:
\begin{equation}
V_{eff}(r) = \frac{l(l + 1)g(r)}{r^{2}} + \frac{g(r)g'(r)}{r}.\label{poten}
\end{equation}

Subsequently, we apply the effective potential identified previously to examine the lower bound on the greybody factor in the essence of our BH solution. Concerning the work of Visser \cite{Visser:1998ke} and Boonserm and Visser \cite{Boonserm:2008zg}, the appropriate method to determine this stringent limit is given by:
\begin{equation} \label{bound}
A_g^2 \geq \operatorname{sech}^{2}\left(\frac{1}{2 \omega} \int_{-\infty}^{\infty}\left|V\right| \frac{d r}{g(r)} \right),
\end{equation}
where $A_g^2$ denotes the transmission coefficient $T$ in this context.
    \begin{figure*}[t!]
      	\centering{\includegraphics[scale=0.9]{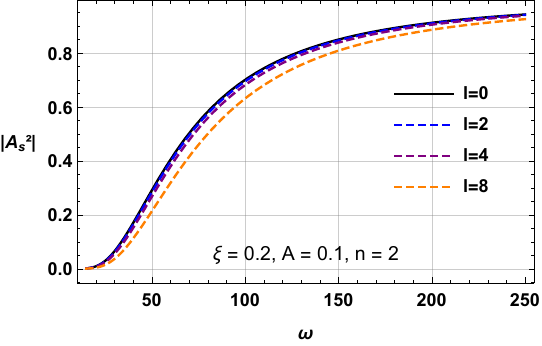}\hspace{5mm}
      	\includegraphics[scale=0.9]{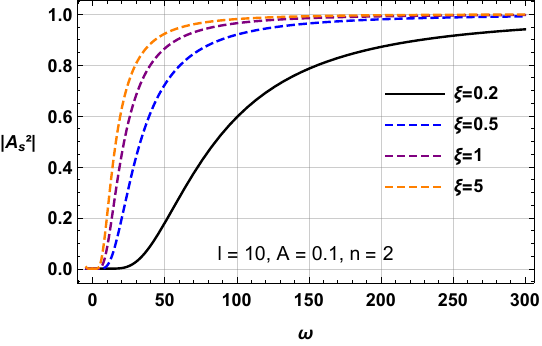} \\
       \includegraphics[scale=0.9]{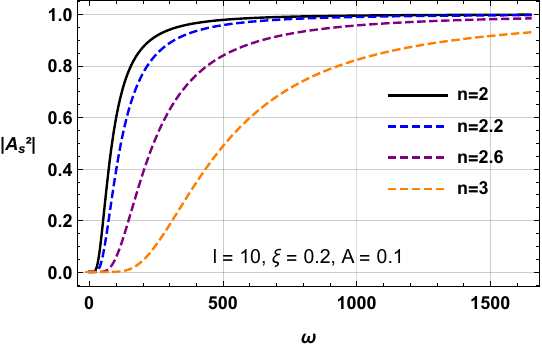}\hspace{5mm}
       \includegraphics[scale=0.9]{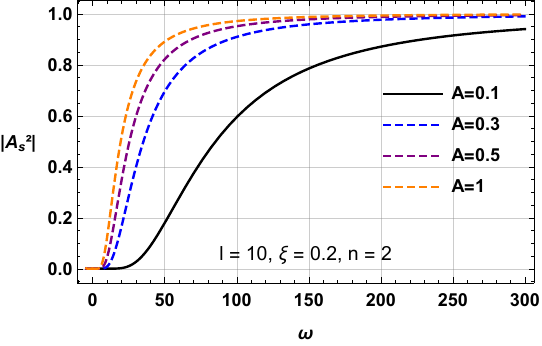}
       
       }
      	\caption{Rigorous bounds on greybody factors of scalar massless field using $M = 1$.}
      	\label{ff9}
      \end{figure*}

In addition, to accommodate the influence of the cosmological horizon, we adjust the boundary conditions as shown by Boonserm et al. \cite{Boonserm:2019mon}. The modified boundary conditions are:
\begin{equation}
A_g^2 \geq A^2_{s}=\operatorname{sech}^{2}\left(\frac{1}{2 \omega} \int_{r_{H}}^{R_{H}} \frac{|V_{eff}|}{g(r)} d r\right)=\operatorname{sech}^{2}\left(\frac{A_{l}}{2 \omega}\right),
\end{equation}
where we define
 \begin{equation}
A_{l}=\int_{r_{H}}^{R_{H}} \frac{|V|}{g(r)} d r=\int_{r_{H}}^{R_{H}}\left|\frac{l(l+1)}{r^{2}}+\frac{g^{\prime}(r)}{r}\right| d r.
\end{equation}

Here, $r_H$ is the event horizon and $R_H$ is the cosmological horizon of the BH. This specification supplies a rigorous lower bound on the grey-body factors relative to the BH solution.

To extract relevant discussion on how the greybody bound varies in relation to the parameters variation, Fig. \ref{ff9} provides insight clarification to this end. It is observed that the variation of the multipole moments involves less variation of the bounds in the sense that it decreases when $l$ increases. Similarly, the variation of the polytropic index leads to an evident and observed variation of the grey body bound where it increases with the decreasing of the polytropic index $n$. On the other hand, the variation of the parameters $\xi$ and $A$ leads to a similar observation in which the bound observed raises whenever the two polytropic parameters $\xi$ and $A$ increase.

\section{BH shadow silhouettes with EHT constraints}\label{Opt}
\begin{figure}[hbt!]
	\centering 
	\includegraphics[width=.49\textwidth]{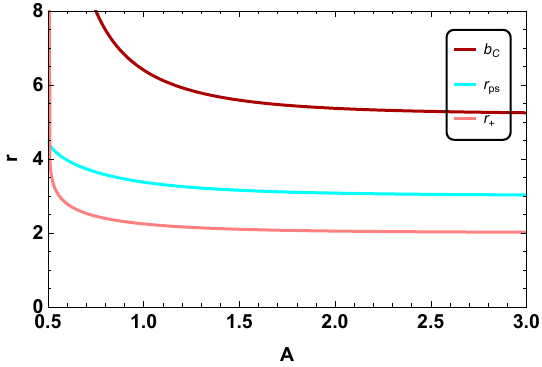}
	\hfill
	\includegraphics[width=.49\textwidth]{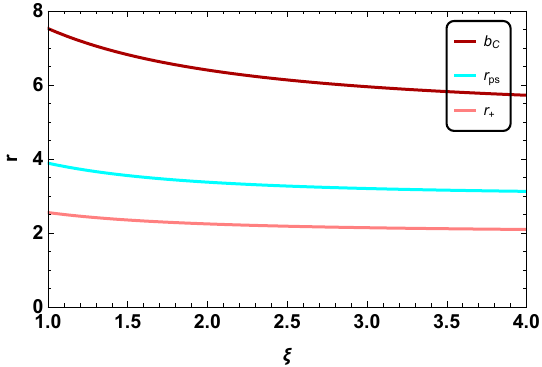}
	\caption{\label{Forbits} 
The plot illustrates the variation of the event horizon, photon sphere radii, and critical impact parameter as functions of the parameters $A$ and $\xi$.	}
\end{figure}

Here, we will examine the shadow radius behavior of a Schwarzschild BH surrounded by a polytropic scalar field gas. Using the observational data from the Event Horizon Telescope (EHT) collaboration for Sgr A* and M87* \cite{EHTApJL2019L1,EHTApJL2019L2,EHTApJL2019L3,EHTApJL2019L4,EHTApJL2019L5,EHTApJL2019L6,EHTApJL2022L12,EHTApJL2022L17,DoSci2019,GRAVITY2022,GRAVITY2020,KocherlakotaPRD2021,VagnozziCQG2023}, we will then put constraints on the model parameters $(A, \xi)$, as reported in Tables \ref{Table:rshxi} and \ref{Table:rshA}. The BH shadow boundary, located on a distant observer's plane, manifests the image of the photon region by separating capture orbits from scattering orbits. As for the photon region itself, it essentially marks the edge of the spacetime region that, if spherically symmetric, corresponds to the photon sphere \cite{PerlickPR2022,CunhaGRG2018,KhodadiPRD2022}. For more convenience, our analysis of null orbits will focus exclusively on a constant polar angle, $ \theta = \pi/2 $. Employing the BH metric from Eq. (3) and the lapse function $g(r)$ defined in Eq. (17), we derive the null geodesics confined to the equatorial plane using the Lagrangian $2\mathcal{L}(x,\dot{x})=g_{\mu \nu }\dot{x%
}^{\mu }\dot{x}^{\nu }$ for the geodesics of a spherically symmetric static spacetime metric, where $	2\mathcal{L}(x,\dot{x}) =( -F(r)\dot{t}^{2}+G(r)\dot{r}^{2} +H(r)\dot{\phi}^{2})$, in which we take $F(r) = g(r)$, $G(r) = 1/F(r)$ and $H(r) = r^{2}$.
Here, the dot above represents the derivative with respect to the affine parameter $\lambda$. By applying the variational principle, the two constants of motion, energy $E$ and angular momentum $L$, can be derived as $E=F(r)\dot{t}$ and $L=H(r)\dot{\phi}$ \cite{ZarePLB2024,ZareJCAP2024,CapozzielloJCAP2023,Capozziello2024,FilhoJCAP2024}.
The impact parameter, which plays a crucial role in the analysis of orbital trajectories, is defined as $b\equiv \frac{L}{E} = \frac{H(r)}{A(r)}\frac{d\phi}{dt}$. Given the metric, setting $ds^2 = 0$ for null geodesics enables us to determine how $r$ varies with $\phi$. In other words, this leads to the derivation of the orbit equation for photon as \cite{EslamPanahEPJC2024,LambiaseEPJC2023,PantigEPJC2022}
\begin{equation}\label{OrbitEq1}
\left(\frac{dr}{d\phi}\right)^{2} = \frac{H(r)}{G(r)}	\left(\frac{h(r)^{2}}{b^{2}}-1\right),
\end{equation}
with $h(r)^2 = \frac{H(r)}{F(r)}$, defined in \cite{PerlickPR2022,Misner}. The function $h(r)$ is particularly useful, as it allows us to determine the location of the photon sphere radius, $r_{ps}$, by setting $h'(r) = 0$, where the prime indicates differentiation with respect to $r$. Applying this condition, we obtain:
\begin{equation}\label{rps}
6M - 2r - r^3 \left(1 + \frac{A^2 r^{\frac{3}{n}}}{\xi}\right)^{-n} \xi^{-n} 
+ r^3 \xi^{-n} \, _2F_1\left(n, n, 1 + n, -\frac{A^2 r^{\frac{3}{n}}}{\xi}\right)=0.
\end{equation}
The photon sphere radius, $r_{ps}$, can only be determined through numerical analysis. As shown in the equation above, it depends on the parameters $n$, $A$, and $\xi$. This is significant because the shadow silhouette and the behavior of the shadow radius are directly influenced by $r_{ps}$, with the critical impact parameter being evaluated at $r_{ps}$.
As a cross-check, it can be shown that for the standard Schwarzschild metric, the photon sphere radius is $r_{ps} = 3M$, and the shadow radius is $R_{sh} = 3\sqrt{3}M$. It is important to highlight that this shadow radius is identical to the critical impact parameter, $b_c$.
To determine the shadow radius $R_{sh}$ as seen by an observer at a distance $r_o$, it is typical to use the angle $\alpha_{sh}$, which represents the angle between the light ray and the radial direction as 
\begin{equation}\label{angle1}
\cot \alpha_{sh} = \left.\sqrt{\frac{G(r)}{H(r)}} \, \frac{d r}{d\phi}\right|_{r = r_{o}} ,
\end{equation}
by applying the appropriate trigonometric identities and the orbit equation, Eq. \eqref{angle1} can be rewritten in an alternative form as
\begin{equation}\label{angle2}
	\sin^{2} \alpha_{sh} = \frac{b_{c}^{2}F(r_{o})}{H(r_{o})}
\end{equation}
Here, $b_c$ is directly linked to the photon sphere radius and can be obtained by satisfying the condition $\frac{d^2 r}{d\phi^2} = 0$, and assuming $G(r) = 1/F(r)$ in this scenario. Consequently, we arrive at:
\begin{equation}\label{bcrit}
b_{c}^{2} = \left.\frac{4 r^{2}}{2F(r)+r F'(r)}\right|_{r = r_{ps}}.
\end{equation}
By employing the lapse function $F(r)$ from Eq. (17) and solving Eq. \eqref{rps} numerically and also considering Eq. \eqref{bcrit} to Eq. \eqref{angle2}, the shadow radius of the BH as observed by a static observer at $r_o$ is given by $R_{sh} = r_o \sin \alpha_{sh} = \sqrt{r_{ps}^{2}F(r_{o})/F(r_{ps})}$. For a static observer located at an asymptotically far distance, this simplifies to 
\begin{equation}\label{shadow1}
R_{sh} = \frac{r_{ps}}{\sqrt{F(r_{ps})}},
\end{equation}
as in the limit $r_o \to \infty$, $F(r_o) \to 1$.
Using numerical plots based on Eqs. \eqref{bcrit}, \eqref{rps}, and the condition $g(r) = 0$, Fig. \ref{Forbits} shows the locations of the event horizon $r_{+}$, photon sphere radii, and critical impact parameter for the corresponding BH solution as functions of the parameters $A$ and $\xi$. Generally, it is observed that all these quantities decrease as the model parameters increase.

\begin{figure}[hbt!]
	\centering 
	\includegraphics[width=.325\textwidth]{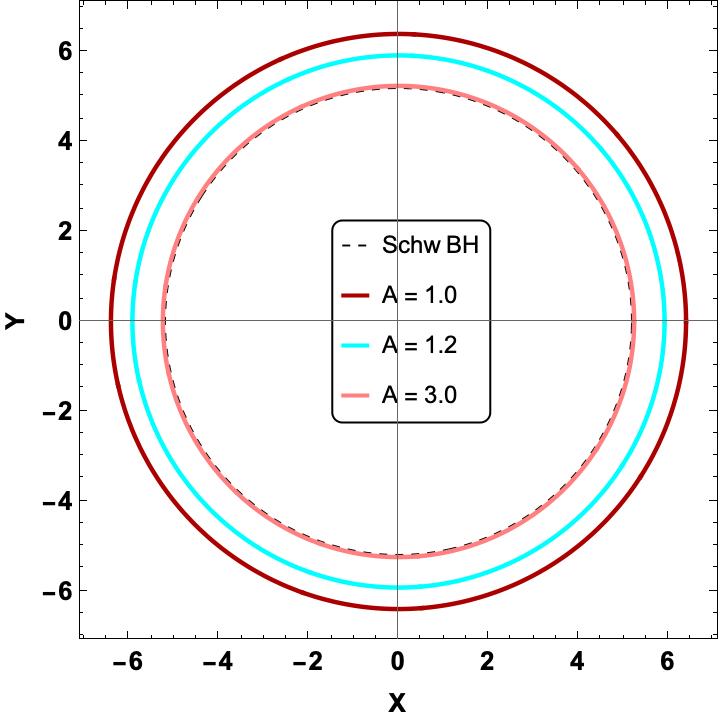}
	\hfill
	\includegraphics[width=.325\textwidth]{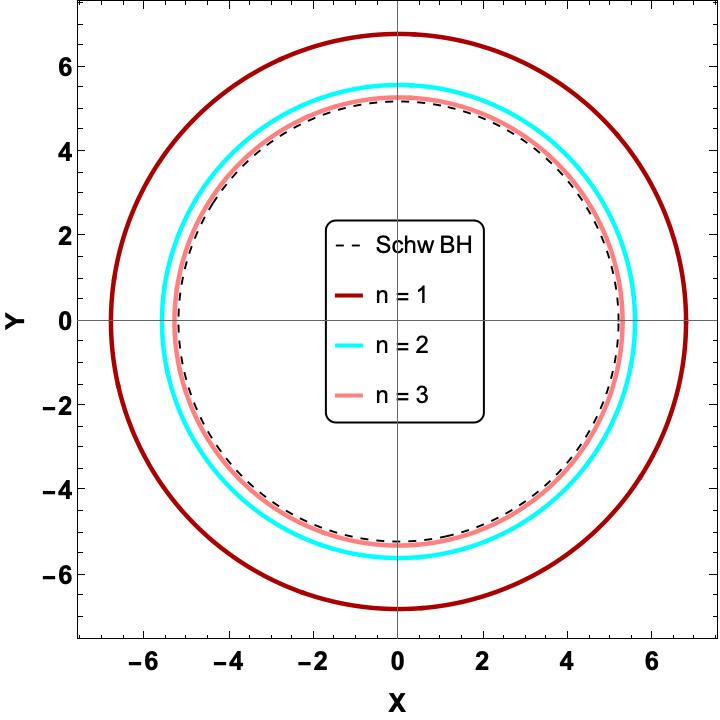}
	\hfill
	\includegraphics[width=.325\textwidth]{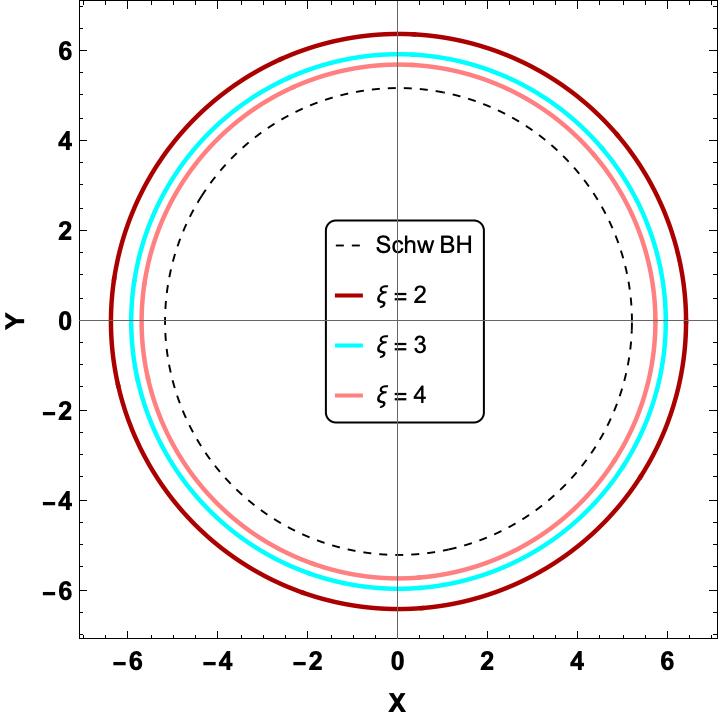}
	\caption{\label{FSh} 
Shadow silhouettes of the BH solution, observed from spatial infinity, for different values of $A$ (left panel), $n$ (middle panel), and $\xi$ (right panel). For comparison, the dashed black curve represents the Schwarzschild BH shadow.}
\end{figure}

In Fig. \ref{FSh}, we display the BH shadow silhouettes for the given BH solution, as observed by an observer at spatial infinity. It is evident that the size of the shadows shrinks as the model parameters $A$, $\xi$, and $n$ increase. This behavior is consistent with expectations, as the critical impact parameter, which defines the shadow radius (and here $R_{sh} = b_c$ in asymptotically flat spacetime), reduces when the model parameters are increased (cf. Fig. \ref{Forbits}).

\begin{figure}[hbt!]
	\centering 
	\includegraphics[width=.49\textwidth]{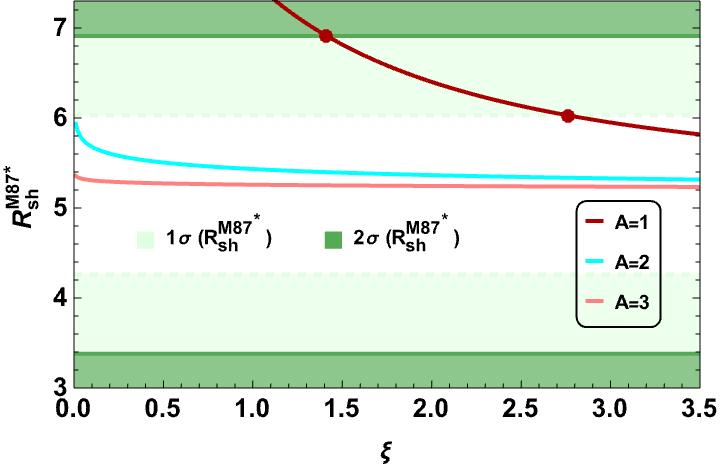}
	\hfill
	\includegraphics[width=.49\textwidth]{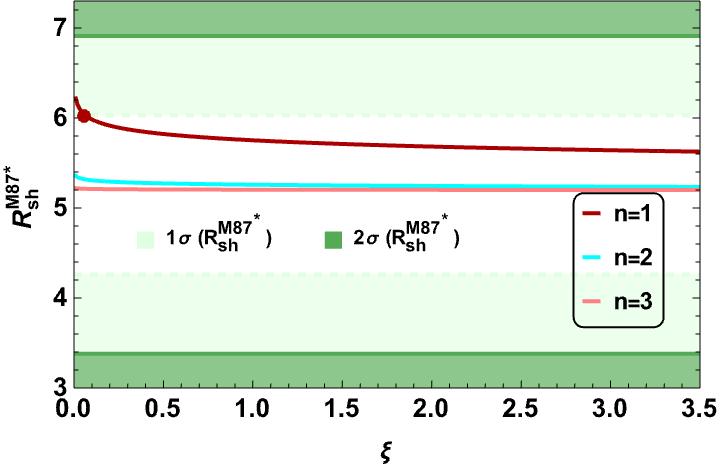}
	\hfill
	\includegraphics[width=.49\textwidth]{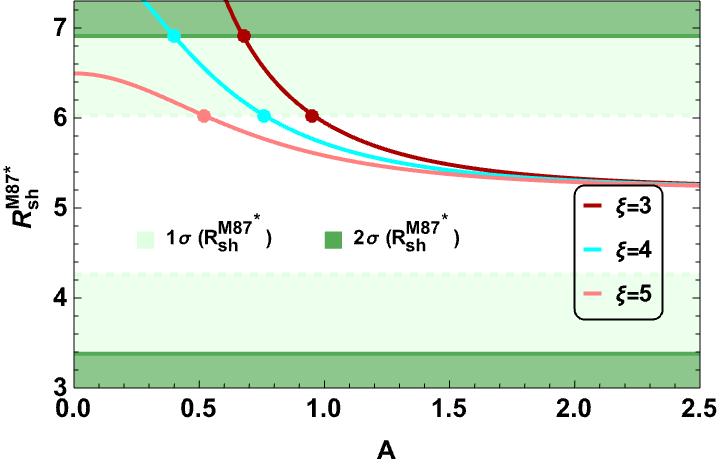}
	\hfill
	\includegraphics[width=.49\textwidth]{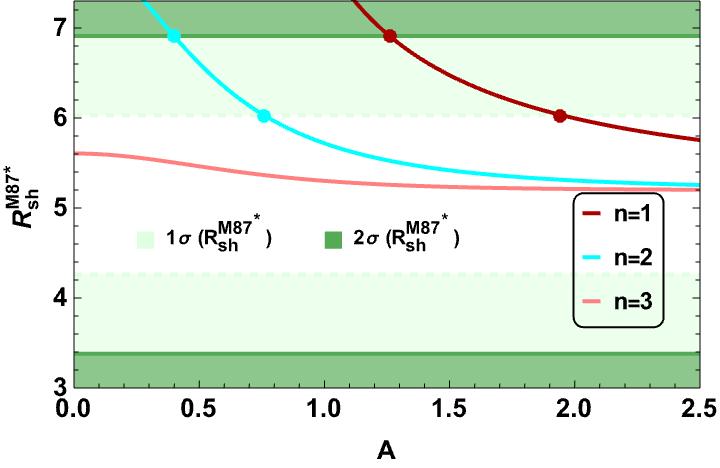}
	\caption{\label{FShConstM87} 
		Top row: Shadow radius shown as a function of $\xi$, evaluated for three different values of $n$ and $A$. Bottom row: Shadow radius plotted as a function of $A$ for three distinct values of $n$ and $\xi$. In both rows, the constraints on the shadow radius are derived using the M87* observations. The white and light green shaded regions correspond to consistency with the EHT image of M87* at the $1\sigma$ and $2\sigma$ CLs, respectively. 
	}
\end{figure}

\begin{figure}[htb]
	\centering 
	\includegraphics[width=.49\textwidth]{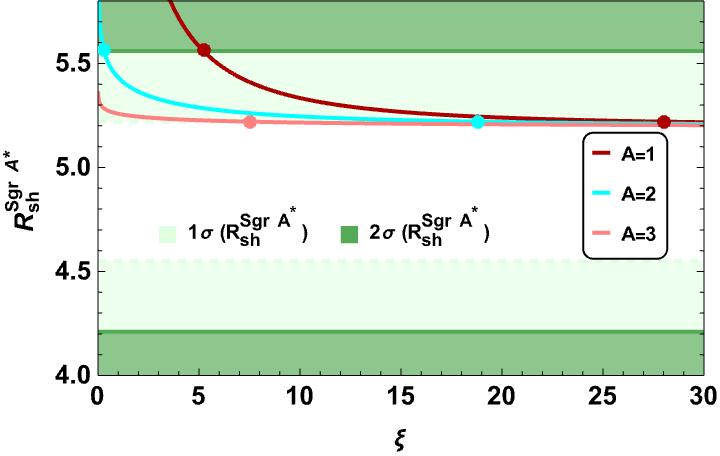}
	\hfill
	\includegraphics[width=.49\textwidth]{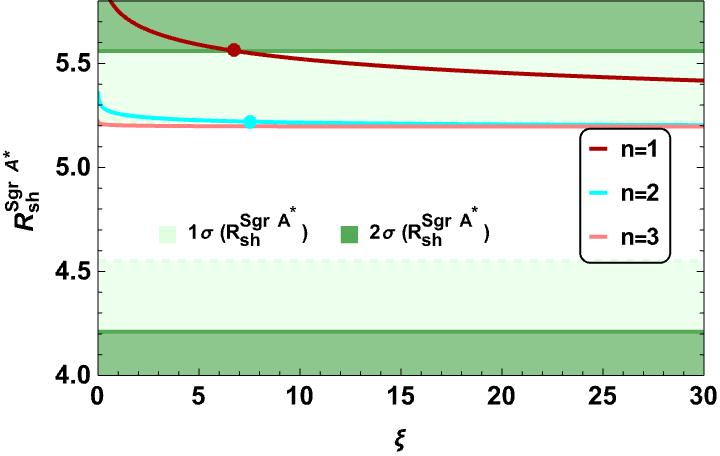}
	\hfill
	\includegraphics[width=.49\textwidth]{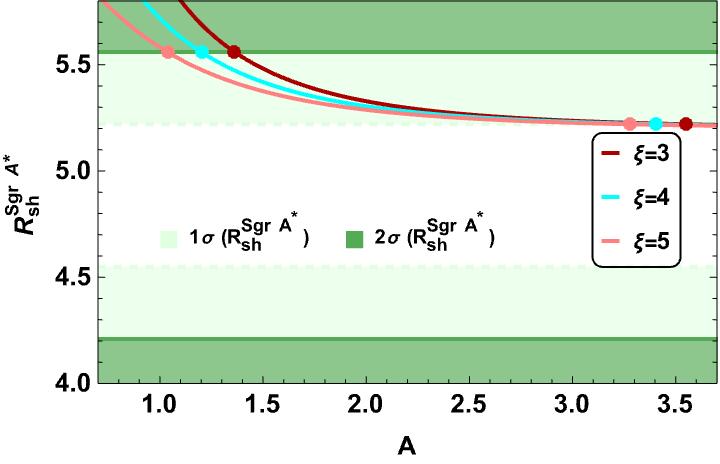}
	\hfill
	\includegraphics[width=.49\textwidth]{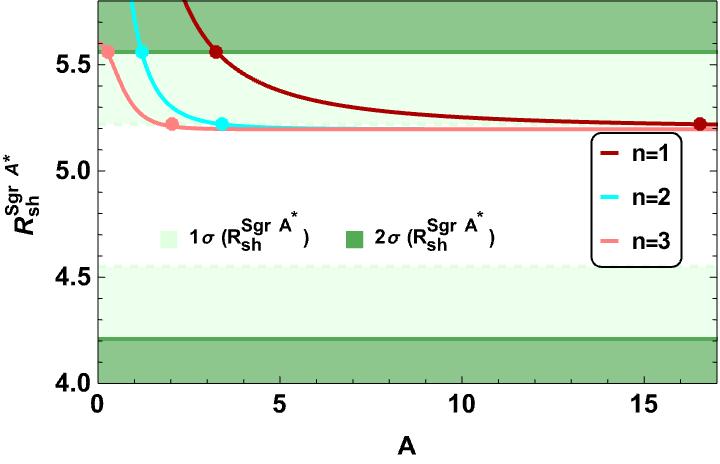}
	\caption{\label{FShConstSgrA} 
		Top row: The shadow radius with respect to $\xi$ for three distinct values of $n$ and $A$. Bottom row: The shadow radius with respect to $A$ for three different values of $n$ and $\xi$.
The constraints derived from the Sgr A* observations are applied to the shadow radius in both rows. The white and light green shaded regions represent the \(1\sigma\) and \(2\sigma\) CLs, respectively, based on the EHT image of Sgr A*.
	}
\end{figure}

The BH shadow observations proposed by the EHT Collaboration have opened an exciting avenue for performing precise tests of gravitational theory in the strong-field and relativistic regimes. Furthermore, incorporating the Schwarzschild deviation parameter $\delta$ offers a powerful tool for placing constraints on the parameters of specific BH models \footnote{For more information on how to determine the $1\sigma$ and $2\sigma$ confidence levels, interested individuals can refer to Refs. \cite{KocherlakotaPRD2021,VagnozziCQG2023} and the re
ferences cited therein.}.
At this point, we aim to put constraints on the model parameters  $A$ and $\xi$ by incorporating the uncertainties provided in Refs. \cite{KocherlakotaPRD2021,VagnozziCQG2023}, along with the shadow image data for M87* and Sgr A* presented by the EHT, as depicted in Figs. \ref{FShConstM87} and \ref{FShConstSgrA}. 

According to Ref. \cite{EHTApJL2019L1}, the M87* BH has a mass of $ M_{\text{M87*}} = (6.5 \pm 0.7) \times 10^9 \, \text{M}_{\odot} $, an angular shadow diameter of $ \theta_{\text{M87*}} = 42 \pm 3 \, \mu\text{as}$, and is located at a distance of $D_{\text{M87*}} = 16.8 \pm 0.8 \, \text{Mpc}$ from Earth. Taking into account the Schwarzschild shadow deviations, $ \delta_{\text{M87*}} = -0.01 \pm 0.17$, where the relation $\frac{R_{\text{S}}}{M} = 3\sqrt{3}(1 + \delta_{\text{M87*}})$ defines the shadow radius levels, the shadow size of M87* is constrained within the range [4.26, 6.03] at the $ 1\sigma $ confidence level (CL).
In this way, the EHT collaboration \cite{EHTApJL2022L12} reports an angular shadow diameter for Sgr A* of $\theta_{\text{Sgr A*}} = 48.7 \pm 7 \, \mu\text{as}$. The inferred distance from Earth to Sgr A* is given as $D_{\text{Sgr A*}} = 8277 \pm 9 \pm 33 \, \text{pc} \, (\text{VLTI})$ and $ 7953 \pm 50 \pm 32 \, \text{pc} \, (\text{Keck})$. Additionally, the BH mass is found as $M_{\text{Sgr A*}} = (4.297\pm 0.012 \pm 0.040)\times10^{6} \text{M}_\odot \, \text{(VLTI)},\,$$ (3.951\pm 0.047) \times 10^{6}\text{M}_\odot \, \text{(Keck)}, \, (4.0^{1.1}_{-0.6})\times10^{6}\text{M}_\odot \, \text{(EHT)}$. 

Based on the Keck and VLTI observations, the fractional deviation from the Schwarzschild shadow expectation for Sgr A* has been quantified as $ \delta_{\text{Sgr A*}} = -0.08^{+0.09}_{-0.09} \, (\text{VLTI}) $ and $ \delta_{\text{Sgr A*}} = -0.04^{+0.09}_{-0.10} \, (\text{Keck}) $. Averaging these values yields an estimate of $ \delta_{\text{Sgr A*}} \simeq -0.06^{+0.065}_{-0.065} \, (\text{Avg}) $. Utilizing the relationship $ \frac{R_{\text{S}}}{M} = 3\sqrt{3}(1+\delta_{\text{Sgr A*}}) $, which governs the shadow radius in terms of the fractional deviation, the shadow size of Sgr A* is consequently restricted to the range $[4.55, 5.22]$ within the $1\sigma$ confidence interval.
We are interested in using these derived constraints to limit the deviation of our BH in question from the standard Schwarzschild solution, specifically by examining how its characteristics diverge from the Schwarzschild case.

As summarized in Tables \ref{Table:rshxi} and \ref{Table:rshA}, the permissible shadow radius range, obtained from the EHT observations of M87* and Sgr A*, imposes constraints on the model parameters $\xi$ and $A$. These constraints are determined at the $1\sigma$ and $2\sigma$ confidence levels, establishing lower bounds for each parameter by independently varying one while keeping the others fixed. 
By modeling the Schwarzschild BH surrounded by a polytropic scalar field gas, as in the cases of the supermassive BHs Sgr A* and M87*, we find that the constraints derived from Sgr A* are more stringent than those from M87*. As shown in Table \ref{Table:rshxi}, when considering a fixed value of $n = 2$ and varying $A$ between $1$ and $3$, the lower bound on $\xi$ ranges from $\xi^{\text{min}} \in (7.56, 28.03)$ at $1\sigma$ CL. while, with $A = 3$ fixed and $n$ varying between $1$ and $3$, the lower bound for $\xi$ remains at $\xi^{\text{min}} = 7.56$ when $n = 2$, also at $1\sigma$ CL.
On the other hand, as detailed in Table \ref{Table:rshA}, the results from Sgr A* reveal that fixing $n = 2$ and varying $\xi$ within the range $3$ to $5$ yields lower bounds for $A$, with $A^{\text{min}}$ spanning $(2.28, 3.55)$ at the $1\sigma$ CL. Meanwhile, holding $\xi = 4$ constant and allowing $n$ to vary between $1$ and $3$ produces a lower bound for $A^{\text{min}}$ ranging from $(2.02, 16.54)$, within the $1\sigma$ confidence interval as well.
Our analysis reveals that the BH parameters are consistent with the EHT observations, indicating that both Sgr A* and M87* could plausibly be modeled as Schwarzschild-like BHs -- characterized here as a Schwarzschild BH surrounded by a polytropic scalar field gas -- within the current precision of astrophysical observations. Consequently, this BH model emerges as a compelling candidate for describing astrophysical BHs.

\begin{table}[ht!]
	\caption{The allowed range of parameter $\xi$, based on the EHT images of
Sgr A* and M87*, with $M=1$.}
	\centering
	\label{Table:rshxi}
	\begin{tabular}{|c|c|c|c|c|}
		\hline
		& \multicolumn{2}{c|}{Sgr A*} & \multicolumn{2}{c|}{M 87*}\\
		\cline{2-5}
		$n=2$ & $1\sigma$ & $2\sigma$ & $1\sigma$ & $2\sigma$ \\
		\cline{2-5}
		\hline
		$A=1$ & $\;(28.03,\,-)\;$ & $\;(5.21,\,-)\;$ & $\;(2.76,\,-)\;$ & $\;(1.41,\,-)\;$\\
		\hline
		$A=2$ & $\;(18.83,\,-)\;$ & $\;(0.30,\,-)\;$ & $\;(-,\,-)\;$ & $\;(-,\,-)\;$\\
		\hline
		$A=3$ & $\;(7.56,\,-)\;$ & $\;(-,\,-)\;$ & $\;(-,\,-)\;$ & $\;(-,\,-)\;$\\
		\hline
		\hline
		\hline
		$A=3$ & $1\sigma$ & $2\sigma$ & $1\sigma$ & $2\sigma$ \\
		\cline{2-5}
		\hline
		$n=1$ & $\;(-,\,-)\;$ & $\;(6.77,\,-)\;$ & $\;(0.06,\,-)\;$ & $\;(-,\,-)\;$\\
		\hline
		$n=2$ & $\;(7.56,\,-)\;$ & $\;(-,\,-)\;$ & $\;(-,\,-)\;$ & $\;(-,\,-)\;$\\
		\hline
		$n=3$ & $\;(-,\,-)\;$ & $\;(-,\,-)\;$ & $\;(-,\,-)\;$ & $\;(-,\,-)\;$\\
		\hline
	\end{tabular}
\end{table}

\begin{table}[ht!]
	\caption{The permissible range of parameter $A$, based on the EHT images of Sgr A* and M87*, with $M=1$.}
	\centering
	\label{Table:rshA}
	\begin{tabular}{|c|c|c|c|c|}
		\hline
		& \multicolumn{2}{c|}{Sgr A*} & \multicolumn{2}{c|}{M 87*}\\
		\cline{2-5}
		$n=2$ & $1\sigma$ & $2\sigma$ & $1\sigma$ & $2\sigma$ \\
		\cline{2-5}
		\hline
		$\xi=3$ & $\;(3.55,\,-)\;$ & $\;(1.36,\,-)\;$ & $\;(0.95,\,-)\;$ & $\;(0.67,\,-)\;$\\
		\hline
		$\xi=4$ & $\;(3.40,\,-)\;$ & $\;(1.20,\,-)\;$ & $\;(0.76,\,-)\;$ & $\;(0.40,\,-)\;$\\
		\hline
		$\xi=5$ & $\;(3.28,\,-)\;$ & $\;(1.04,\,-)\;$ & $\;(0.52,\,-)\;$ & $\;(-,\,-)\;$\\
		\hline
		\hline
		\hline
		$\xi=4$ & $1\sigma$ & $2\sigma$ & $1\sigma$ & $2\sigma$ \\
		\cline{2-5}
		\hline
		$n=1$ & $\;(16.54,\,-)\;$ & $\;(3.26,\,-)\;$ & $\;(0.95,\,-)\;$ & $\;(0.68,\,-)\;$\\
		\hline
		$n=2$ & $\;(3.40,\,-)\;$ & $\;(1.20,\,-)\;$ & $\;(0.76,\,-)\;$ & $\;(0.40,\,-)\;$\\
		\hline
		$n=3$ & $\;(2.02,\,-)\;$ & $\;(0.25,\,-)\;$ & $\;(0.52,\,-)\;$ & $\;(-,\,-)\;$\\
		\hline
	\end{tabular}
\end{table}


\section{Emission rate} \label{EM}
By examining the shadow of the BH, we can look at particle emissions in the neighborhood of the BH solution. It has been shown that, for a far-away observer, the BH shadow is an estimate of the BH absorption cross-section at a high-energy limit \cite{EE1}. Broadly speaking, in the case of a spherically symmetric topology of a BH, the absorption cross-section exhibits oscillatory behavior around a constant threshold value of $\sigma_{lim}$ at very high energies. As the shadow provides a visible indication of a BH, it is broadly estimated to be equivalent to the surface area of the photon sphere, which can be reckoned at $\sigma_{lim} \approx \pi R_{s}^{2}$. The energy emission rate is defined in the following way \cite{EE1,EE2,EE3,EE4}:
\begin{equation}\label{Eqemission}
\frac{d^{2}E(\varpi )}{dtd\varpi }=\frac{2\pi ^{3}\varpi 
^{3}R_{s}^{2}}{e^{\frac{\varpi }{T_H}}-1}, 
\end{equation}%
in which $\varpi $ is the emission frequency, and $T_H$ is the Hawking temperature.
\begin{figure*}[hbt!]
\centering{
\includegraphics[scale=0.9]{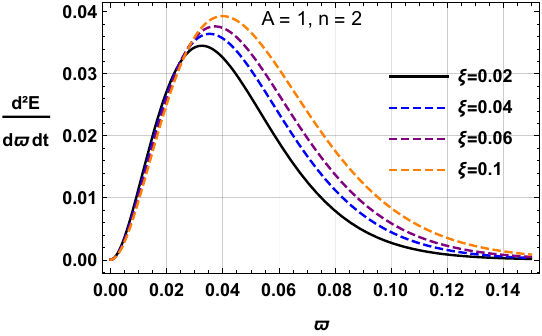}
\hspace{0.2cm}
\includegraphics[scale=0.9]{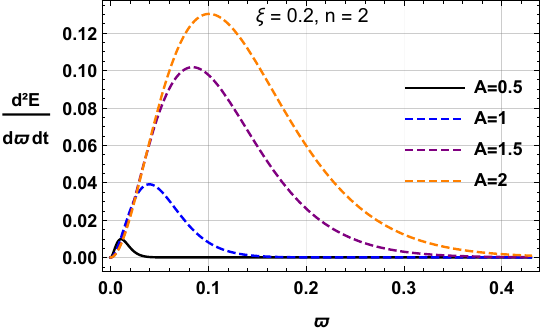}\\
       \includegraphics[scale=0.9]{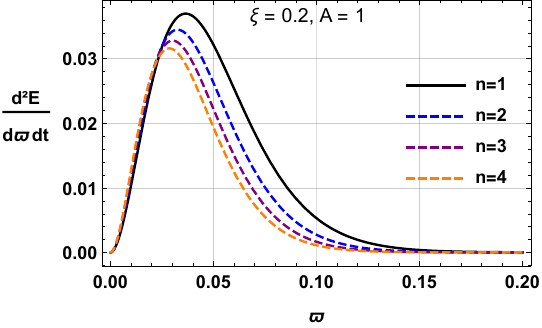}
}
\caption{Emission rate of the black hole  with respect to $\varpi$ using $M = 1$ for various value of the parameter space.}\label{ff14}
\end{figure*}

By exploiting the Hawking temperature formula exposed in Eq. \eqref{TTT}, we have illustrated the variation of the  emission rate concerning $\varpi$ for various values of $\xi$, $A$  and $n$ in Fig. \ref{ff14}. It is noticeable that as the polytropic index $n$ increases, the emission rate of the BH diminishes. Conversely, the increasing in the parameter $\xi$ and $A$, the emission rate of the BH increases. So, the analysis demonstrates that as the polytropic index $n$ raises, the BH’s evaporation rate decreases, and its lifespan extends. In contrast, when the polytropic parameters such that $\xi$ and $A$ grow, the black hole’s evaporation rate gradually increases, and its lifespan shortens.

\section{Conclusion}\label{Con}
In this study, we examined the QNMs, greybody factors, shadow behavior, and sparsity process of BHs with a surrounding polytropic scalar field gas. The results showed that all the parameters of the polytropic structure, as well as the multipole moment, have a consistent impact on the variation of the QNMs, in particular the frequency oscillations and the damping rate of the GWs. In practice, the impact of parameter $A$ on the frequency oscillations and the damping rate is much more significant than $\xi$. Evidently, as $A$ and $\xi$ increase, the frequency oscillation and the damping rate of GWs raise. Also, the same behavior is observed for the variation of the multipole moment as it increases, causing an increase in the oscillation frequency and the damping rate of GWs. Specifically, changes in the frequencies and damping rates of the QNMs exert a dominant impact on the typical ‘ringdown’ signal emitted by merging black holes. Variations in the parameters assigned to the polytropic scalar field surrounding the black hole typically trigger changes in the effective potential, which ultimately influence the oscillation frequencies and decay times of the QNMs. 

In GWs observations, these changes might point to the existence of fluid matter surrounding black holes. In principle, detectors like LIGO and Virgo, which are ringdown sensors, can discriminate between BHs in the ordinary vacuum and those surrounded by a gaseous dark energy model in the form of a polytropic structure. Nevertheless, to ensure that the parameters of the model are constrained with the required precision, we might need to hold off until the development of space-based GWs detectors like LISA. Subtle discrepancies in QNM frequencies or damping rates can offer valuable insights into the possibility of the presence of polytropic structure or even point to changes in GR in the proximity of BHs. This possibility raises new prospects for uncovering the fundamental character of black hole surroundings using GWs astronomy.

Our findings concerning the greybody factors based on a rigorous analytical approach revealed interesting results in favor of exploring the impact carried out by the polytropic structure parameters together with the multipole moment $l$. Small values of the multipole moments $ l $ and the polytropic index $n$ lead to much larger bounds for the grey body. Whereas the increase of parameters $\xi$ and $A$ causes a rise in the bound of the greybody factor. 

We found that the lower bounds on the model parameters derived from M87* observations are smaller than those obtained from Sgr A*. Additionally, the constraint on $A$ is significantly more stringent than that on $\xi$ for both M87* and Sgr A*. Notably, the Sgr A* data impose tighter limits on the model parameters, as they extend beyond the upper $1\sigma$ CLs. Consequently, within the parameter space consistent with our model, the EHT observations do not rule out the existence of a surrounding polytropic scalar field gas at galactic centers. This report offers one of the first constraints on the polytropic scalar field gas based on EHT data from both M87* and Sgr A*.
We also looked at the emission rate of the BH with a surrounding polytropic structure. Related observations showed that as the polytropic index $n$ increases, the evaporation rate of the black hole diminishes. On the other hand, a growth in the value of $\xi$ and $A$ raises the evaporation rate, implying a reduction in the lifetime of the BH.

\section*{Acknowledgments}
The research of L.M.N. and S.Z. was supported by the Q-CAYLE project, funded by the European Union-Next Generation UE/MICIU/Plan de Recuperacion, Transformacion y Resiliencia/Junta de Castilla y Leon (PRTRC17.11), and also by RED2022-134301-T and PID2023-148409NB-I00, financed by MI-CIU/AEI/10.13039/501100011033. The research of H.H. was partially supported by the Long-Term Conceptual Development of a University of Hradec Kr\'{a}lov\'{e} for 2023, issued by The Ministry of Education, Youth and Sports of the Czech Republic.

\end{document}